\documentclass[aip,graphicx]{revtex4-1}

\usepackage{booktabs}       
\usepackage{amsfonts}       
\usepackage{amsmath}
\usepackage{nicefrac}       
\usepackage{microtype}      
\usepackage{float}          
\usepackage{graphicx}
\usepackage{subcaption}     
\usepackage[export]{adjustbox}
\usepackage{wrapfig}
\usepackage{siunitx}        
\graphicspath{{fig/}}       
\usepackage{amsmath}        
\usepackage{multirow}       
\usepackage{xcolor}

\draft

\begin{document}

\title{Operator learning for predicting multiscale bubble growth dynamics}

\author{Chensen~Lin}
\affiliation{Division of Applied Mathematics, Brown University, Providence, RI 02912, USA}

\author{Zhen~Li}
\affiliation{Department of Mechanical Engineering, Clemson University, Clemson, SC 29634, USA}

\author{Lu~Lu}
\affiliation{Department of Mathematics, Massachusetts Institute of Technology, Cambridge, MA 02139, USA}

\author{Shengze~Cai}
\affiliation{Division of Applied Mathematics, Brown University, Providence, RI 02912, USA}

\author{Martin~Maxey}
\affiliation{Division of Applied Mathematics, Brown University, Providence, RI 02912, USA}

\author{George~Em~Karniadakis}
\thanks{Corresponding author. E-mail: \texttt{george\_karniadakis@brown.edu}}
\affiliation{Division of Applied Mathematics, Brown University, Providence, RI 02912, USA}

\date{\today}

\begin{abstract}
Simulating and predicting multiscale problems that couple multiple physics and dynamics across many orders of spatiotemporal scales is a great challenge that has not been investigated systematically by deep neural networks (DNNs). 
Herein, we develop a framework based on operator regression, the so-called deep operator network (DeepONet), with the long term objective to simplify multiscale modeling by avoiding the fragile and time-consuming ``hand-shaking" interface algorithms for stitching together heterogeneous descriptions of multiscale phenomena. To this end, as a first step, we investigate if a DeepONet can learn the dynamics of different scale regimes, one at the deterministic  macroscale and the other at the stochastic microscale regime with inherent thermal fluctuations.  
Specifically, we test the effectiveness and accuracy of DeepONet in predicting multirate bubble growth dynamics, which is described by a Rayleigh-Plesset (R-P) equation at the macroscale and modeled as a stochastic nucleation and cavitation process at the microscale by dissipative particle dynamics (DPD). 
First, we generate data using the R-P equation for multirate bubble growth dynamics caused by randomly time-varying liquid pressures drawn from Gaussian random fields (GRF).  Our results show that properly trained DeepONets can accurately predict the macroscale bubble growth dynamics and can outperform long short-term memory (LSTM) networks. We also demonstrate that DeepONet can extrapolate accurately outside the input distribution using only very few new measurements.
Subsequently, we train the DeepONet with DPD data corresponding to stochastic bubble growth dynamics. Although the DPD data is noisy and we only collect sparse data points on the trajectories, the trained DeepONet model is able to predict accurately the mean bubble dynamics for time-varying GRF pressures.
Taken together, our findings demonstrate that DeepONets can be employed to unify the macroscale and microscale models of the multirate bubble growth problem, hence providing new insight into the role of operator regression via DNNs in tackling realistic multiscale problems and in simplifying modeling with heterogeneous descriptions.
\end{abstract}

\pacs{}

\maketitle

\section{Introduction}

Machine learning methods that learn effective physical models from available observational data have attracted increasing attention in recent years~\cite{rudy2017data,champion2019data,lu2019deeponet} and have demonstrated success in diverse application areas~\cite{raissi2020hidden,raissi2019physics,jin2020sympnets,chen2020physics}.
In particular, the use of deep neural networks (DNNs) as a surrogate model of complex dynamic systems is a promising approach for exploring hidden physics from observed data~\cite{raissi2020hidden,raissi2019physics}. Many different DNN architectures have been developed for targeting different types of scientific problems~\cite{carleo2019machine,cichy2019deep}. Examples include convolutional neural networks (CNN) for turbulence modeling~\cite{lapeyre2019training}, recurrent neural networks (RNN) for predictive flow control~\cite{bieker2020deep}, and generative adversarial networks (GAN) for super-resolution of fluid flows~\cite{xie2018TempoGan}. 
In general, different physical quantities involved in the same dynamic process are correlated either explicitly or implicitly. Using the observational data of one physical quantity to predict other unobserved quantities requires a model to correctly correlate different physical quantities, which can be achieved by training a DNN with a lot of observational data~\cite{peng2020multiscale}.   
There are two possible scenarios when we develop a DNN model for a physical problem. First, if we have sufficient physical knowledge on the dynamic process and know its governing equations, i.e., in the form of partial differential equations (PDEs), it becomes possible to encode the PDEs into the DNN, which leads to the physics-informed neural network (PINN) model~\cite{raissi2019physics}. Second, if we lack physical knowledge of a given physical problem and do not know the form of its governing equations, we need to learn the governing equations or operators from available observed data. This motivated Lu et al.~\cite{lu2019deeponet} to develop a deep operator network (DeepONet) to learn implicitly PDEs from data as well as many other explicit and implicit diverse operators. 

Both PINN and DeepONet models have been successfully applied to mostly single scale problems, including high-speed flows~\cite{mao2020physics}, hidden fluid mechanics~\cite{raissi2020hidden}, and diffusion-reaction processes~\cite{lu2019deeponet}.
However, it is still challenging to apply DNNs to truly multiscale problems described by different mathematical formulations.  The form of governing equations for various scales could be significantly different even for the same physical problem, which makes the integration and coupling of heterogeneous physical models in multiscale problems very difficult. Because the physical models constructed by DeepONet do not have an explicit form of governing equations, we can possibly develop a unified multiscale framework based on DeepONet to significantly simplify the multiscale modeling procedure. In the present work, we will investigate this possibility by considering multiscale bubble growth dynamics, described by the Rayleigh-Plesset (R-P) equation at the continuum scale and modelled as a stochastic nucleation and cavitation process at the microscale. 

Bubbles are present in a wide range of applications, from advanced materials~\cite{yu2017superwettability} to biology and medicine~\cite{dollet2019bubble}, as either laser-generated or acoustically-driven bubbles~\cite{jamburidze2017high}. The heat transfer and fluid flow processes associated with liquid-vapor phase change phenomena are among the most complex transport conditions that are encountered in engineering applications. The complexities typically arise from the interplay of multiple length and time scales involved in the process, non-equilibrium effects, and other effects 
associated with dynamic interfaces. Nucleation of one phase into another occurs at the atomic scales whereas growth of the nuclei and their interaction with the system occurs at much longer length and time scales~\cite{novak2007molecular}. For example, boiling is a complex nonequilibrium multiscale process, where the physics is linked from nano to meso to macro scales. For each of these processes, the physics from different scales have to be seamlessly connected. 
However, different models are valid or are practical only within some space and time range, and bridging the gap between models of different scales has been a challenge for decades.

A simple and fundamental problem to understand bubble dynamics is that of the growth of a single bubble as the pressure changes in an otherwise quiescent fluid. 
This problem has a long history, beginning with the important contributions of Rayleigh~\cite{rayleigh1917viii}. In general, contemporary interest in the problem is motivated largely by the sound produced by bubbles that undergo a time-dependent change of volume, particularly in the context of cavitation bubbles that are produced in hydromachinery or in the propulsion systems of submarines and other underwater vehicles or for biomedical applications. 
The objective for single-bubble dynamics is usually to predict the bubble radius as a function of time in a preset ambient pressure field. Once we know the bubble radius as a function of time, we can determine the velocity and pressure fields throughout the liquid, and evaluate the sound produced. 

In the macroscale regime, the R-P ordinary differential equation (ODE) is a valid model for the governing dynamics of a spherical bubble in an incompressible fluid. It is derived from the Navier–Stokes (N-S) equation based on the assumptions of a continuum hypothesis and of a single spherically symmetric bubble in an infinite domain of liquid at rest far from the bubble and with uniform temperature far from the bubble. 
In the microscale and nanoscale regimes, we can use  
molecular dynamics (MD) models to investigate the single bubble nucleation problem~\cite{maruyama1997molecular}. For longer term development,
dissipative particle dynamics (DPD) is a more efficient coarse-grained MD method that has been used to model complex fluid systems such as red blood cells~\cite{li2013continuum}, polymers~\cite{wang2020controlled}, droplet~\cite{li2013three, zhang2019self}, suspension~\cite{lin2020dissipative}, etc. 
The equations of DPD are stochastic differential equations with conservative, dissipative and random forces.
A few studies have reported using DPD to simulate bubble dynamics~\cite{wu2017sliding, tran2013rheology}. An effective  model~\cite{pan2018mesoscopic} is a hybrid model of DPD and its extension to multi-body dissipative particle dynamics (MDPD), where the traditional DPD particles are used for gas phase and MDPD particles for liquid phase. The model is validated by comparing it to the R-P equation for the larger size bubbles and with MD for the smaller size bubbles.

In this paper, our goal is to test the effectiveness of DeepONet as the surrogate model for both the macroscale and microscale physics. 
The paper is organized as follows. In Section 2, we describe details of the DeepONet model. In Section 3, we introduce the macroscopic model (R-P model) for bubble dynamics and evaluate the performance of DeepONet at the continuum scale. In Section 4, we introduce the microscopic model (DPD model) and evaluate the performance of DeepONet at the stochastic microscale. Finally, we conclude the paper in Section 5.

\section{Learning operators via DeepONet}  \label{sec.ONet}

Our work is inspired by the work of Lu et al.~\cite{lu2019deeponet}, who proposed a deep operator network (DeepONet) to approximate nonlinear continuous operators. The DeepONet architecture is based on rigorous mathematical theory, namely the {\it universal operator approximation theorem}~\cite{chen1995universal}, which has been extended to DNNs.

\begin{figure}[htbp]
\centering
\begin{subfigure}{0.9\textwidth}
\includegraphics[width=0.9\linewidth]{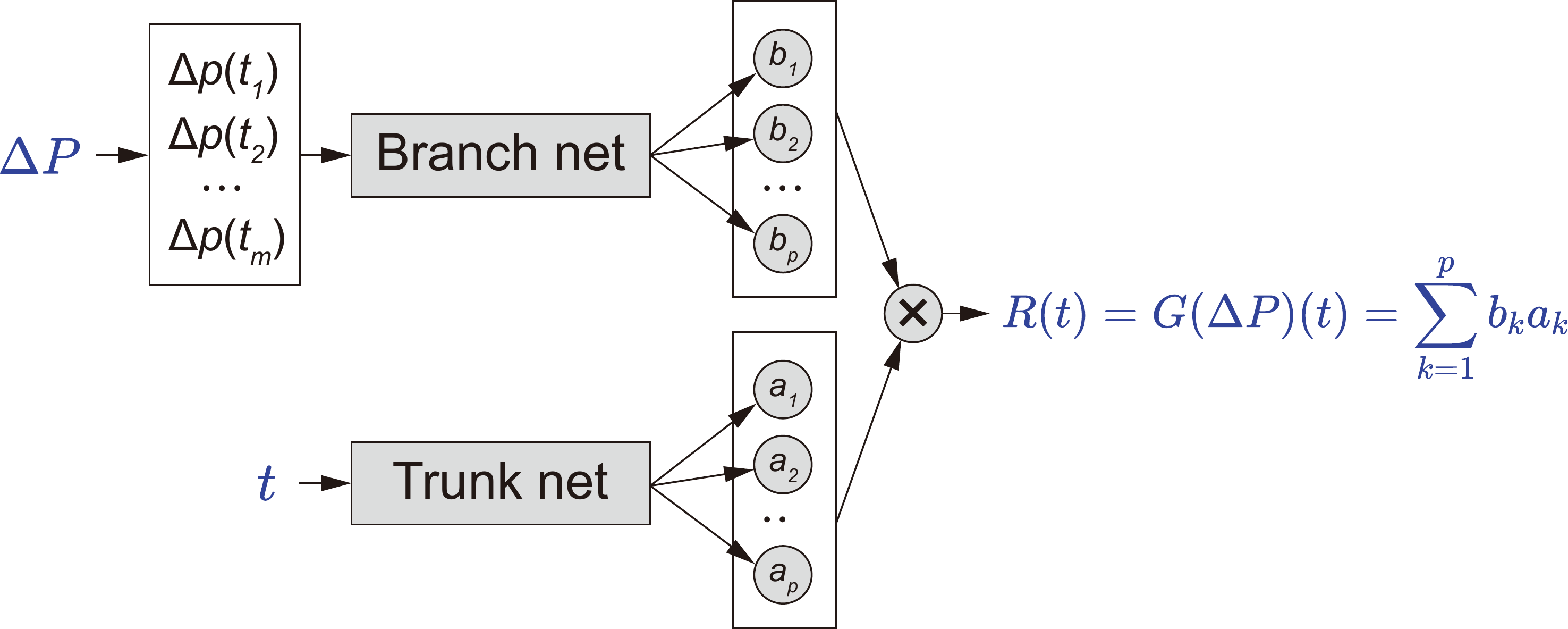}
\end{subfigure}
\caption{Schematic of the DeepONet architecture. The branch network takes the liquid pressure change at different times (from $t_1$ to $t_m$) as input and outputs $[b_1,b_2,...,b_p]^T$. The trunk network takes time $t$ as input and outputs $[a_1,a_2,...,a_p]^T$. Then, the two vectors are merged together to give the prediction of bubble radius $R(t)=G(\Delta P)(t)=\sum_{k=1}^{p}(b_{k}a_{k})$. Both the branch and trunk are feedforward neural networks (FNNs).}
\label{fig.illu.ONet.net}
\end{figure}

The architecture of the DeepONet consists of a DNN for encoding the discrete input function space (branch net) and another DNN for encoding the coordinates of the output functions (trunk net). 
In the bubble dynamics application, the input function is the ambient liquid pressure change deviating from the initial equilibrium state, $\Delta P$, while the output function is the bubble radius, $R(t)$. 
There are different ways to represent a function but here the input to the branch is represented by discrete function values at a sufficient but finite number of locations, which we call ``sensors". 
The architecture we used in this study is shown in Fig.~\ref{fig.illu.ONet.net}. The trunk network takes time $t$ as the input and outputs $[a_1,a_2,...,a_p]^T$. The branch network takes $[\Delta p(t_1),\Delta p(t_2),...,\Delta p(t_m)]^T$ as the input and outputs a vector $[b_1,b_2,...,b_p]^T$. This architecture corresponds to the unstacked DeepONet~\cite{lu2019deeponet}. Then, the two vectors are merged together to predict the bubble radius: $R(t) = G(\Delta p)(t) = \Sigma^{p}_{k=1}b_{k}a_{k}$.

The loss function we employ is the mean squared error (MSE) between the true value of $R(t)$ from the R-P equation and the network prediction derived from the input $([\Delta p(t_1),\Delta p(t_2),...,\Delta p(t_m)],t)$. 
DeepONet is a high-level network architecture without restricting the architectures of its inner trunk and branch networks. Here, we choose feedforward neural network (FNN) as the architectures of the two sub-networks, which are implemented in the Python library DeepXDE \cite{lu2019deepxde}.

\begin{figure}[htbp]
\centering
\begin{subfigure}{0.5\textwidth}
\includegraphics[width=0.9\linewidth]{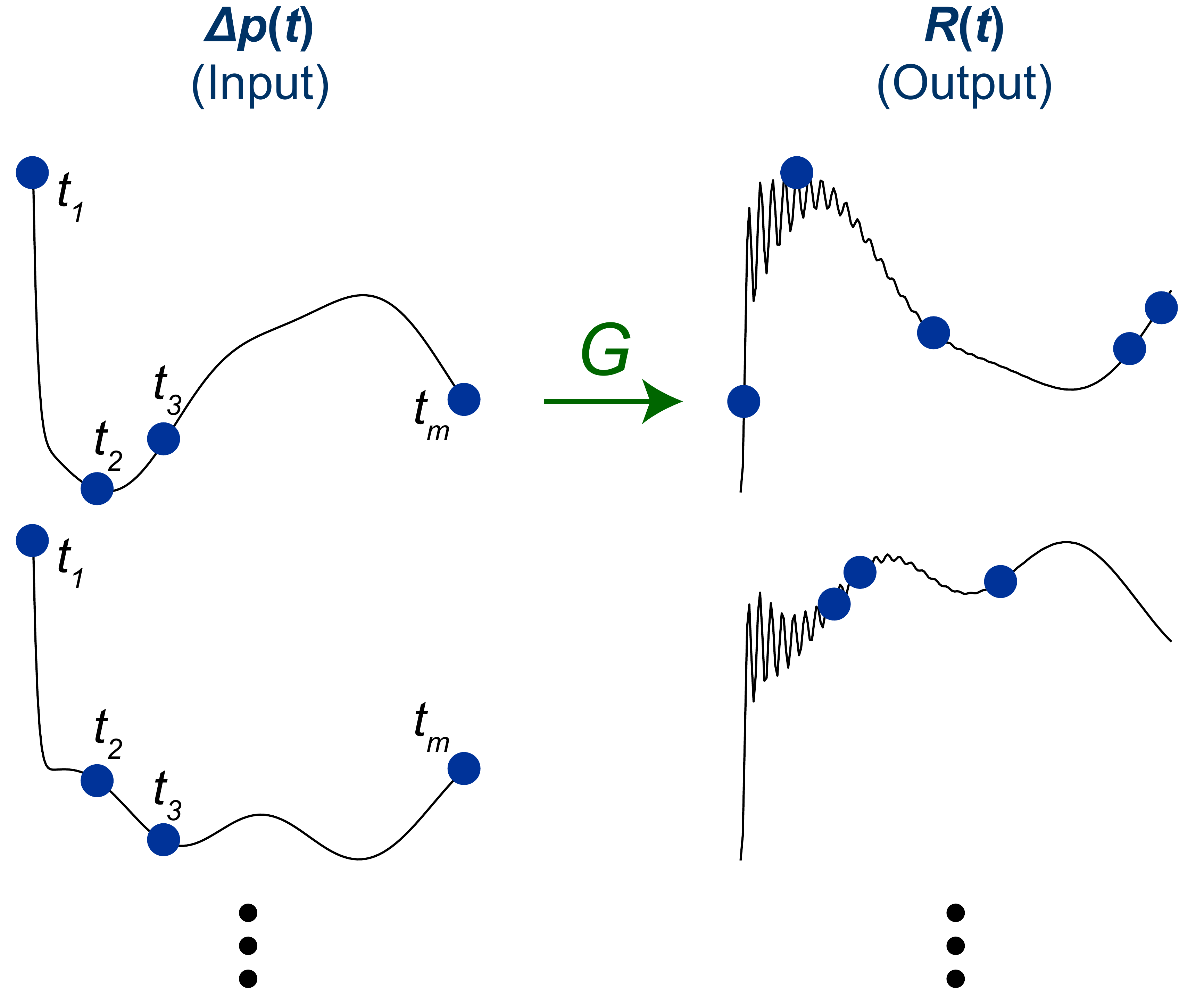} 
\end{subfigure}
\caption{Illustration of the training data for two different trajectories in time. For each input function $\Delta p(t)$, DeepONet requires the evaluations at the same discrete time. However, DeepONet does not have any constraints on the number or location for the evaluation of the output function $R(t)$. Hence, the data points for the output function can be very sparse and randomly spaced.}
\label{fig.illu.ONet.US}
\end{figure}

For each input function $\Delta P$, DeepONet requires evaluations at the same set of discrete times. However, DeepONet does not have any constraint on the number or location for the evaluation of output function $R(t)$. 
DeepONet provides a continuous estimate of the bubble radius, $R$ at any arbitrary time.
In this study, for the efficient training of the networks, we choose a fixed number of evaluations for each $R(t)$ but at randomly selected times within the whole interval; see Fig.~\ref{fig.illu.ONet.US}.

\section{Macroscale (continuum) regime}  \label{sec.RP}
\subsection{Rayleigh–Plesset model}

The R-P equation~\cite{plesset1977bubble, brennen2013cavitation, prosperetti2017vapor} is the classical continuum-level model that describes the change in radius $R(t)$ of a single gas-vapor bubble in an infinite body of liquid as the far-field pressure $P_{\infty}(t)$ varies with time $t$. 
This ODE for the dynamics of a spherical bubble is based on a number of assumptions. Although it is limited to the case where a continuum model is valid, it is a simple yet powerful tool to reveal the effects of the surroundings on bubble response and it is the starting point for our investigation of operator learning as applied to bubble dynamics.

\begin{figure}[htbp]
\centering
\begin{subfigure}{0.6\textwidth}
\includegraphics[width=0.8\linewidth]{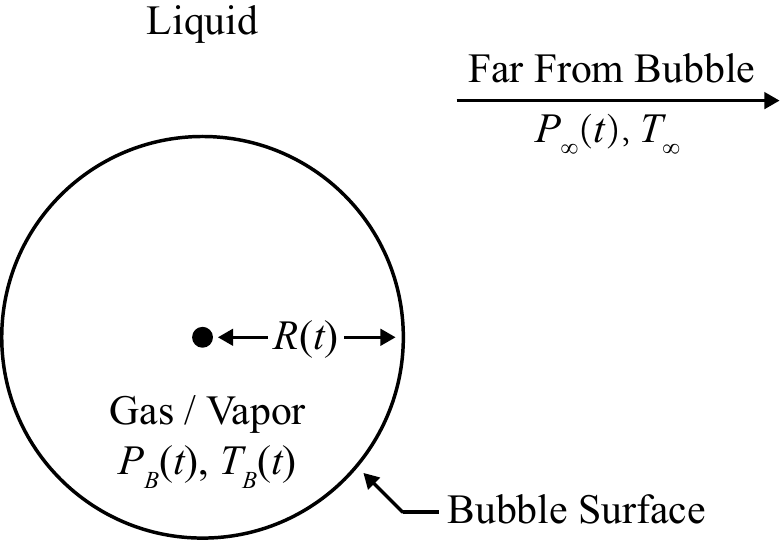}
\end{subfigure}
\caption{Schematic of a spherical bubble in an infinite liquid.}
\label{fig.schematic}
\end{figure}

The underlying assumptions of the R-P model are that the liquid is at rest far from the bubble and that the liquid phase is incompressible, with constant density $\rho_L$ and kinematic viscosity $\nu_L$. It is also assumed that the contents of the bubble have negligible inertia, are homogeneous and that the temperature, $T_{B}(t)$, and pressure, $P_{B}(t)$, within the bubble are uniform. The configuration is illustrated in Fig.~\ref{fig.schematic}. As the bubble expands it generates a radial potential flow corresponding to a point source at the bubble center. The pressure in the liquid can be determined from the Navier-Stokes equations together with the balance of the liquid pressure at the bubble surface with the bubble pressure $P_B$ and the surface tension between the phases, given by the coefficient of surface tension $\gamma$. As the liquid responds to the expanding bubble, the liquid pressure balances the effects of liquid inertia and damping of the motion by viscosity. The R-P equation can be written as

\begin{equation}
{{\frac {P_{B}(t)-P_{\infty}(t)}{\rho _{L}}}
=R{\frac {d^{2}R}{dt^{2}}}+{\frac {3}{2}}\left({\frac {dR}{dt}}\right)^{2}+{\frac {4\nu _{L}}{R}}{\frac {dR}{dt}}+{\frac {2\gamma }{\rho _{L}R}}}.
\label{eq.RP}
\end{equation}

The general R-P model assumes that the bubble contains a mixture of an ideal gas and liquid vapor. The pressure inside the bubble is then the sum of the partial pressures. The vapor pressure $P_{V}$ is determined by the temperature within the bubble $T_{B}$ while the pressure of the gas $P_{G}$ is given by a polytropic gas law,
\begin{equation}
{P_{B}(t) = P_{V}(T_{B}) + P_{G0} \left(\frac{T_B}{T_\infty}\right)
\left(\frac{R_0}{R}\right)^{3k}}.
\label{eq.gas}
\end{equation}
The coefficient $k=1$ corresponds to isothermal conditions, or $k=1.4$ for a simple adiabatic gas.
Since the microcopic simulations to be described later are based on a thermostat to maintain a constant temperature, we will neglect the effects of any temperature difference between the bubble and the liquid, setting $T_{B} = T_{\infty}$, and assume isothermal conditions. With these assumptions the R-P equation becomes,
\begin{equation}  \label{eq.RP.gas}
{{\frac{P_{V}(T_{\infty})-P_{\infty}(t)}{\rho_{L}}} + 
\frac{P_{G0}}{\rho_{L}} \left(\frac{R_0}{R}\right)^{3}
= R{\frac{d^{2}R}{dt^{2}}}+{\frac {3}{2}}\left({\frac {dR}{dt}}\right)^{2}+{\frac{4\nu _{L}}{R}}{\frac{dR}{dt}}+{\frac {2\gamma }{\rho_{L}R}}}.
\end{equation}

The initial conditions imposed here are that the initial microbubble radius is $R_0$ at $t=0$ and that the bubble is in static equilibrium, with the gas pressure $P_{G} = P_{G0}$ and the fluid at a pressure $P_{\infty}(0)$ so that
\begin{equation}  \label{eq.initial.condition}
	{P_{G0}=
		P_{\infty}(0)-P_{V}(T_{\infty})+\frac{2\gamma}{R_{0}}}.
\end{equation}
Consistent with initial equilibrium,  $dR/dt|_{t=0}=0$. Under these conditions the vapor pressure simply gives a constant offset in the liquid pressure at infinity. So as to reduce the number of parameters under consideration, the results here are based on $P_{\infty}(0)=P_{V}(T_{\infty})$ as might occur close to the onset of cavitation. We are not trying to model any specific physical set of conditions but for convenience then the initial gas pressure is given in terms of the surface tension, $P_{G0}= 2\gamma/R_{0}$.

\begin{figure}[htbp]
\begin{subfigure}{0.9\textwidth}
\includegraphics[width=0.9\linewidth]{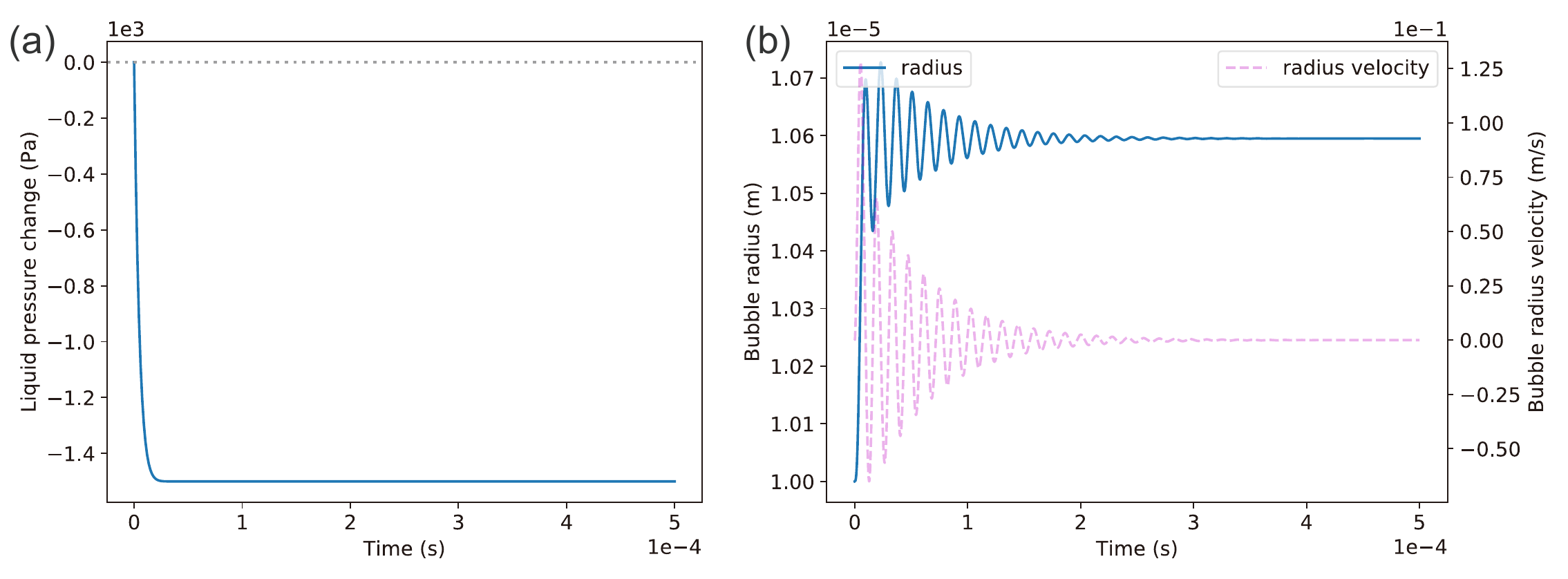}
\end{subfigure}
\caption{Typical solution of the R-P equation as a bubble responds to a rapid negative change in pressure. (a) Ambient liquid pressure change; (b) Bubble radius (blue) and radius velocity (pink).}
\label{fig.typical}
\end{figure}

A typical solution for Eq.~\eqref{eq.RP.gas} under these conditions is shown in Fig.~\ref{fig.typical} with a sharp change in the pressure $P_{\infty}(t)$, which drops sharply below $P_{\infty}(0)$ to a new constant value. The pressure drop induces a series of damped oscillations as the bubble first expands rapidly and then gradually adjusts to a new equilibrium size. In the absence of dissipation by viscosity, these oscillations would continue indefinitely without attenuation.  For small variations in the liquid pressure $P_{\infty}(t) = P_{\infty}(0)+\Delta p(t)$, the response is $R(t)=R_{0} +r(t)$ and the linearized form of Eq.~\eqref{eq.RP.gas} is,
\begin{equation}  \label{eq.RP.linear}
- \frac{\Delta p(t)}{\rho_{L}} = R_{0} \frac{d^2 r}{dt^2} + \frac{4 \nu_{L}}{R_{0}} \frac{dr}{dt} + \frac{1}{\rho_{L} R_{0}} \left( 3P_{G0} - 2\frac{\gamma}{R_{0}} \right) r.
\end{equation}
Under the present conditions, the last coefficient is equal to $2P_{G0}$ or equivalently $4 \gamma/R_{0}$. The corresponding natural frequency $\omega_0$ for undamped oscillations is given by
\begin{equation}
\omega_{0}^2 = 2\frac{P_{G0}}{\rho_{L}R_{0}^{2}}.
\end{equation}
This shows the natural time scale of the bubble response to changing pressure and an indication of how often the bubble radius normally should be observed. It is also clear that this time scale is dependent on $R_0$.

For the training and test data to use with DeepONet, we consider a bubble that is subject to a sharp low-pressure at time $t=0$ and then experiences a random liquid pressure fluctuation thereafter. The liquid pressure change comprises of two parts: the random fluctuation generated by Gaussian random fields (GRFs), and the ramp function, given in terms of a power law, for the initial drop in pressure $P_{\infty}(t)$. We define the liquid pressure change, $\Delta P(t)$, as:

\begin{equation}    \label{eq.GRF}
\begin{split}
   & g(t) \sim \mathcal{GP}(\mu, \sigma k_{lT}(t_1, t_2)),        \\
   & s(t) = \begin{cases}
                  \left( \frac{t}{qT} \right)^{10}  &  \quad 0 \leq t < qT,  \\
                  1  &  \quad qT \leq t \leq T,  \\
                  \end{cases}                  \\
   & \Delta P(t) = g(t)s(t).                    \\
\end{split}
\end{equation}
%
The Gaussian covariance kernel $k_{lT}(t_1, t_2) = \exp(-\| t_1 - t_2 \|^2 / 2l^{2}T^{2})$ has a correlation time-scale based on the time-scale parameter $lT > 0$. $T$ is the time interval, and the dimensionless parameter $l$ determines the smoothness of the sampled function. We choose $l = 0.2$ unless otherwise stated. The value of $q$ indicates the duration of the initial pressure drop, and we choose $q=0.1$ unless otherwise stated. The mean value of the pressure change is $\mu = -1500 \si{Pa}$ and the standard deviation is $\sigma = 400 \si{Pa}$.

The setup of the problem is summarized in Table~\ref{table.coeff}. The parameters are representative of water as the liquid phase and the surface tension is set for a water-air interface at room temperature.

\begin{table}[hbp]
\scriptsize
\centering
\caption{Values of parameters in Eq. \ref{eq.RP.gas}.}
\label{table.coeff}
\begin{tabular}{|c|c|c|} 
\hline
Parameter       & Physical meaning        & Value  \\ 
\hline
 ${\rho _{L}}$  & liquid density          & \SI{1e3}{\kilogram\meter^{-3}}       \\ 
\hline
 ${\nu _{L}}$   & viscosity               & \SI{1e-6}{\meter^2\second^{-1}}       \\ 
\hline
 ${\gamma}$     & surface tension         & \SI{7.2e-2}{\newton \meter^{-1}}       \\ 
\hline
 $T$            & time interval               & \SI{5e-4}{\second}       \\ 
\hline
 $R(t_0)$       & initial radius          & \SI{1e-5}{\meter}       \\ 
\hline
 $dR(t_0)/dt$   & initial radius velocity & \SI{0}{\meter/\second}       \\
\hline
\end{tabular}
\end{table}

\begin{figure}[htbp]
\centering
\begin{subfigure}{0.9\textwidth}
\includegraphics[width=1.0\linewidth]{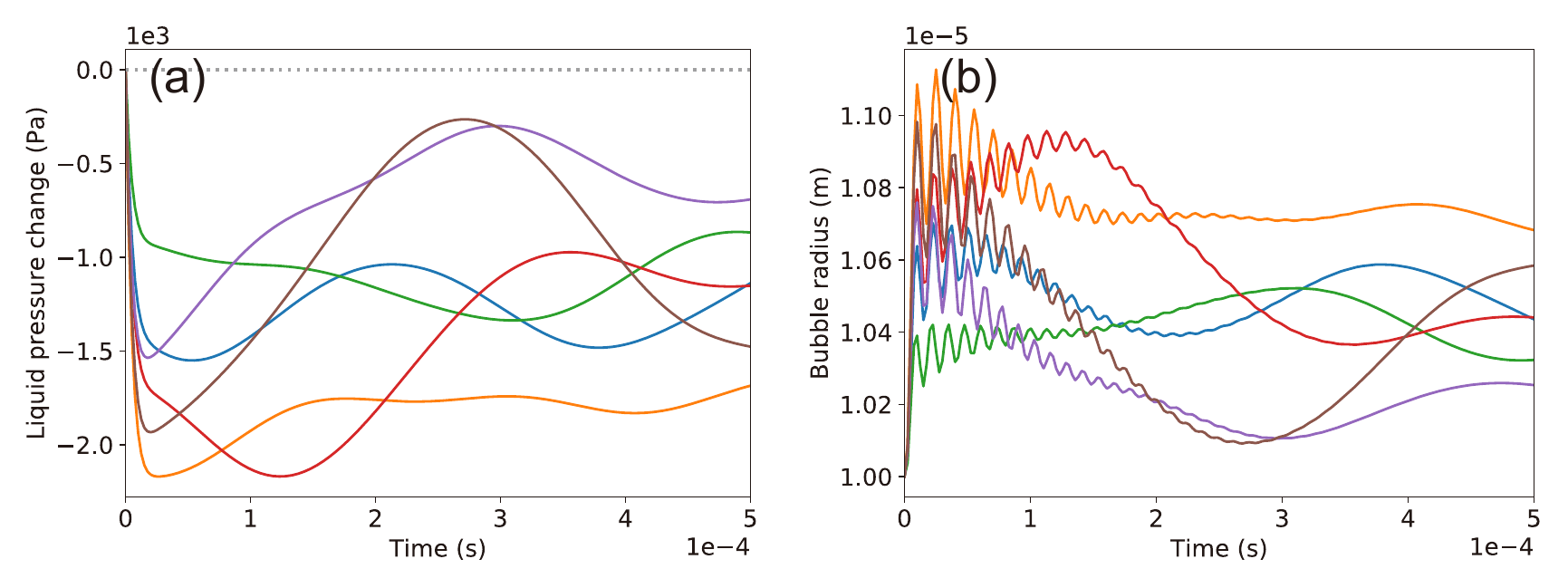}
\end{subfigure}
\caption{(a) Sample trajectories of liquid pressure change with time ($\Delta P(t)$); (b) Corresponding bubble radius ($R(t)$) from training dataset. The corresponding data for $\Delta P(t)$ and $R(t)$ are presented with the same color.}
\label{fig.problem.illu.sameR0}
\end{figure}

Fig.~\ref{fig.problem.illu.sameR0} demonstrates some representative data for the liquid pressure change, $\Delta P(t)$, and corresponding bubble radius, $R(t)$. The general features of these solutions are characteristic of the response of a bubble as it passes through any low-pressure region.

\clearpage
\subsection{Convergence tests}

\begin{figure}[bp]
\begin{subfigure}{1.0\textwidth}
\includegraphics[width=1.0\linewidth]{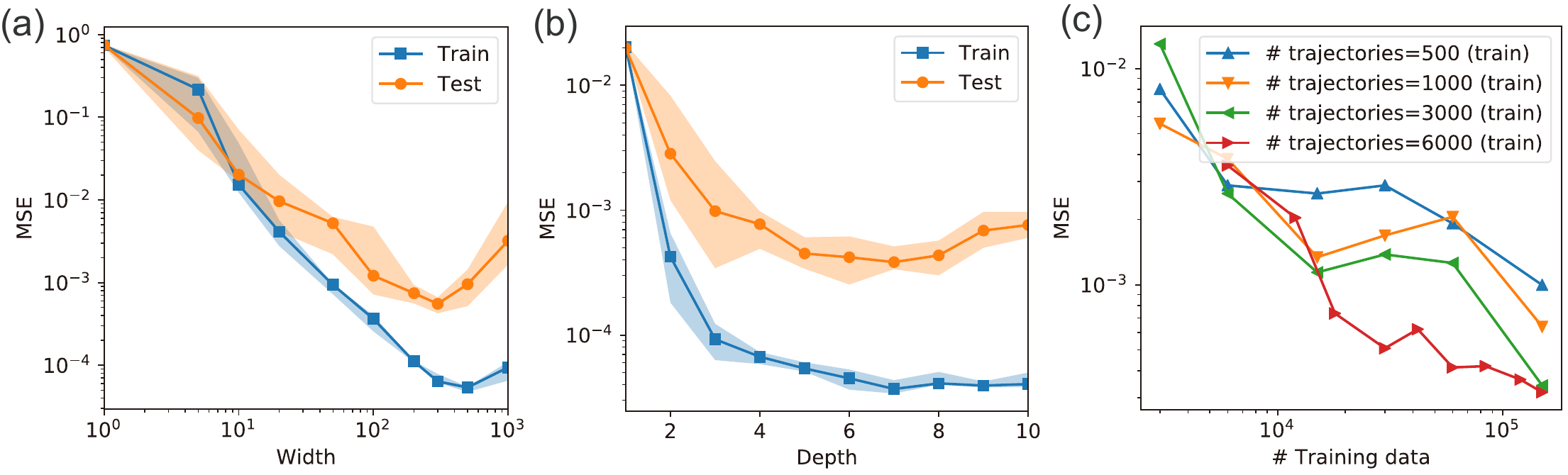}
\end{subfigure}
\caption{Mean square error (MSE) convergence. (a) Training and testing errors versus network width (both branch and trunk net) (depth 2); (b) Training and testing errors versus network depth (both branch and trunk net) (width 100); (c) Testing errors versus size of training data (width 200, depth 3 for both branch and trunk net).}
\label{fig.convergence}
\end{figure}

\begin{figure}[bp]
\centering
\begin{subfigure}{0.9\textwidth}
\includegraphics[width=1.0\linewidth]{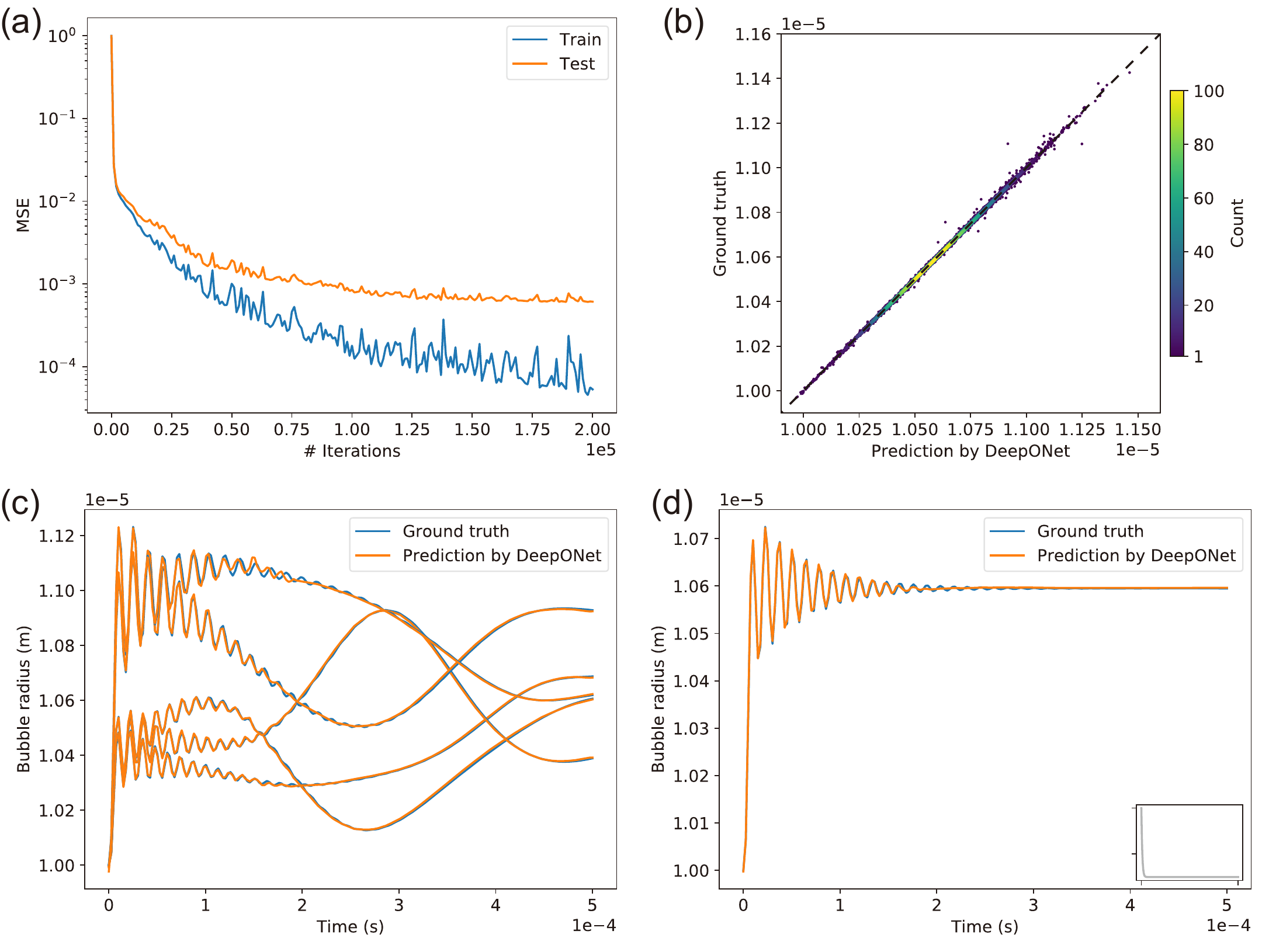}
\end{subfigure}
\caption{Performance of DeepONet for bubble growth prediction. (a) Loss histories of the training  and testing; (b) Parity plot of DeepONet prediction and ground truth from test dataset; (c) Sample predicted $R(t)$ trajectories and ground truth from test dataset (from GRF); (d) Prediction for a pressure field not from GRF, which is a typical solution of R-P equation discussed in Sec.\ref{sec.RP}. The inset plot is the ambient pressure change.}
\label{fig.best}
\end{figure}

Here, we investigate how the network size and training dataset affect the convergence of DeepONet.
We apply the Adam optimizer with a small learning rate $5\times10^{-4}$ to train the DeepONet for $2\times10^{5}$ iterations. The rectified linear unit (ReLU) is used for the activation function.
By varying the network width, we observe that there is an optimal width to achieve the smallest error (depth 2), see Fig.~\ref{fig.best}(a). We note that increasing the width from 1 to 500 could decrease the error, but the error would increase instead when the width further increases.
We also note that increasing the depth from 1 hidden layer to 10 hidden layers could decrease the training error (width 100), but the test loss reaches the minimum when the depth is about 6. 
To examine the effect of the training dataset size, we choose networks of width 200 and depth 3 to eliminate the network size effect, and the test errors using different dataset size are shown in Fig.~\ref{fig.convergence}(c). As DeepONet has no constraint on the number of data points per trajectory, we can get the same number of data points either from more trajectories with sparser points per trajectory or fewer trajectories with denser points per trajectory. 
Fig.~\ref{fig.convergence}(c) shows the influence of the number of trajectories and the size of the dataset. It is reasonable that including more training data points leads to smaller error, and we can also see that when the number of data points is fixed, the number of trajectories is usually more important than the density of data points on each trajectory.

Fig.~\ref{fig.best} shows some representative results of the best model. The width and depth of the network is 200 and 3, respectively, and the number of data points is $1\times10^{5}$ (5000 trajectories). 
The loss histories of the training and testing are presented in Fig.~\ref{fig.best}(a); the losses converge to rather small values. Upon training, the DeepONet can predict the bubble radius at any time when the input functions are given. 
The comparison of prediction values against the ground truth is shown in Fig.~\ref{fig.best}(b), in which the diagonal line indicates a perfect prediction. The points are colored based on the neighbor density because the points near the diagonal line are highly overlapped. 
Some $R(t)$ trajectories predicted by DeepONet are shown in Fig.~\ref{fig.best}(c) along with the ground truth. DeepONet can predict them very accurately, capturing all the oscillations, decay, and smooth change afterward. We also use the trained DeepONet to predict an out-of-distribution pressure input signal (not from GRF), which is the typical solution of R-P equation discussed in Sec. \ref{sec.RP}, see the results in Fig.~\ref{fig.best}(d) (the inset subplot is the pressure field).

\subsection{Comparison of DeepONet and LSTM}
Considering the nature of the present problem where we aim to forecast time series, it is reasonable to question if any recurrent neural network (RNN) could also be a suitable tool for forecasting besides DeepONet.
RNNs were initially used for language processing models, due to their ability to memorize long-term dependencies. However, when time lags increase, gradients of RNNs may vanish through unfolding RNNs into deep FNNs. Due to the gradient vanishing problem, LSTM~\cite{hochreiter1997long} was proposed with forget units, which were designed to give the memory cells the ability to choose when to forget certain information, thus determine the optimal time lags. The LSTM cell contains mainly four parts: input gate, input modulation gate, forget gate and output gate. The input gate takes a new input point from outside and processes newly coming data; the input modulation gate takes input from the output of the LSTM cell in the last iteration; the forget gate decides when to forget the output results and thus selects the optimal time lag for the input sequence; and the output gate takes all results calculated and generates output for the LSTM cell. 
LSTM has been utilized for traffic flow prediction~\cite{fu2016using}, vessel dynamics prediction~\cite{del2019learning}, human trajectory prediction~\cite{alahi2016social}, stock price prediction~\cite{nelson2017stock}, etc. 

\begin{figure}[ht]
\begin{subfigure}{1.0\textwidth}
\includegraphics[width=1.0\linewidth]{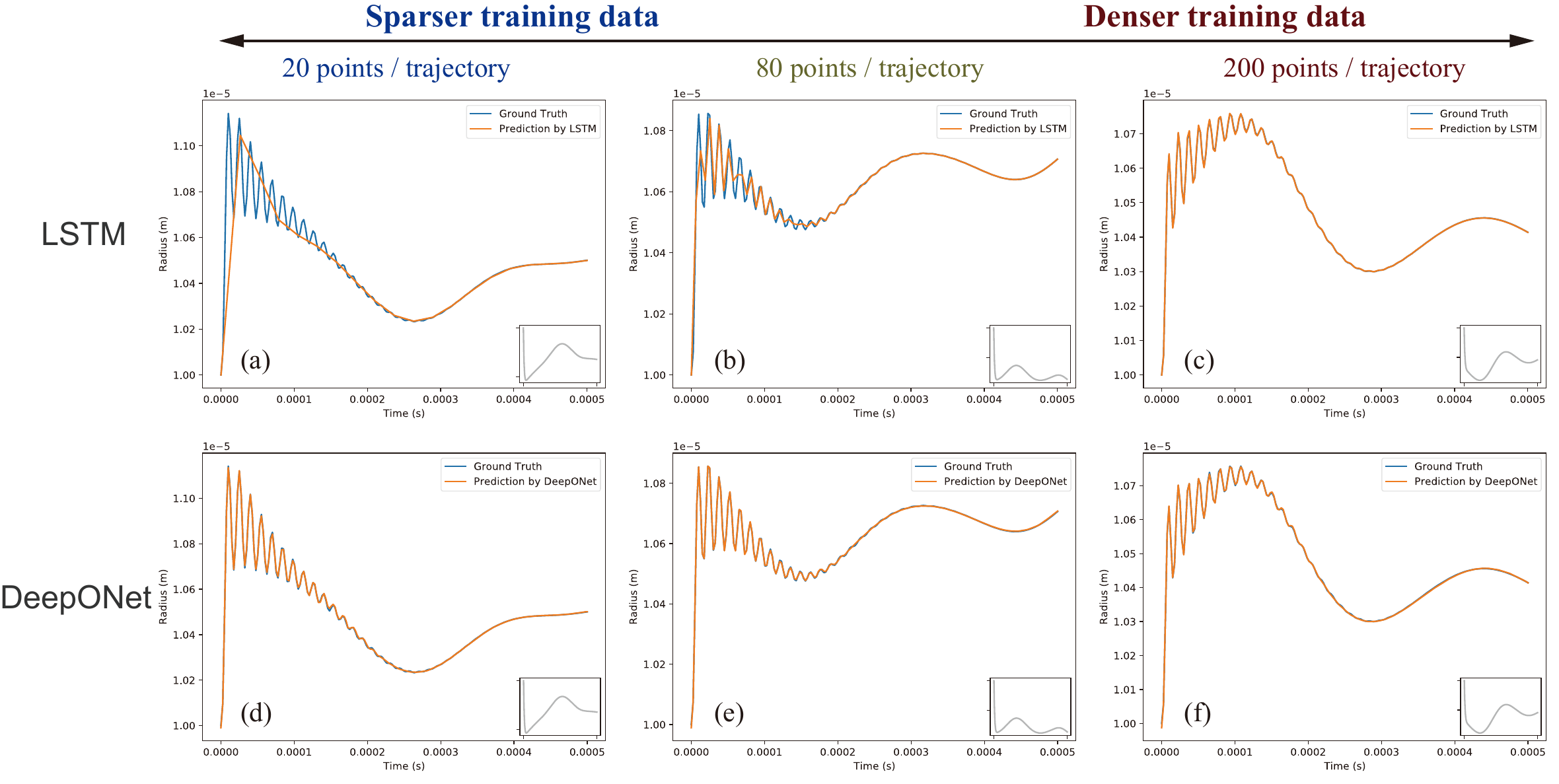}
\end{subfigure}
\caption{Comparative performance of LSTM and DeepONet on testing data. (a-c) Predictions of LSTM; (d-f) Predictions of DeepONet. The number of training data points on each trajectory are (a)(d) 20, (b)(e) 80, (c)(f) 200. The inset subplots show the liquid pressure trajectory. 5000 training trajectories are provided for all the cases. DeepONet show great advantage when training data are very sparse.}
\label{fig.ONet.vs.LSTM}
\end{figure}

The data used to train and test LSTM are the same as those used for DeepONet. With ample training data and tuned hyper-parameters, we find that both DeepONet and LSTM can handle this problem well. 
DeepONet has a great advantage due to its flexibility in handling arbitrary training data points. We develop six benchmark cases and plot the results in Fig.~\ref{fig.ONet.vs.LSTM}. The inset subplots are the input functions. For a fair comparison, the input functions are the same for LSTM and DeepONet. 
To train the model, 5000 pairs of liquid pressure and bubble radius trajectories are provided while the availability of data points on the radius trajectories is limited. In Fig.~\ref{fig.ONet.vs.LSTM}(a)(d), only 20 data points are known per trajectory, (b)(e) 80 data points, and (c)(f) 200 data points. Because LSTM requires that the locations of all data points be aligned, even though LSTM learned well on these given locations, it does not have any knowledge on locations which are not on a predefined grid, see Fig.~\ref{fig.ONet.vs.LSTM}(a)(b). The problem can be alleviated when the data points are sufficiently dense  (Fig.~\ref{fig.ONet.vs.LSTM}(c)).
On the other hand, as stated in Sec.~\ref{sec.ONet}, DeepONet does not enforce any constraint on the number or locations for the evaluation of the output function. That means the 20/80/200 training data points can be not only arbitrary but also at different locations for each trajectory. 
This improvement enables DeepONet to predict accurately based on any arbitrary locations, no matter how sparse the training data are, see Fig.~\ref{fig.ONet.vs.LSTM}(d)(e).
From Fig.~\ref{fig.ONet.vs.LSTM}(d), we note the predicted trajectory does not deviate from the ground truth.
Moreover, from our experience, once the training data is ample, DeepONet can predict smooth and accurate results even if only a single  data point is collected per trajectory.
In general, we have no prior knowledge about the right time interval at which to perform measurements or we may have difficulty in acquiring very dense data, and hence DeepONet shows a clear advantage both for its accuracy and versatility.

\subsection{Predictions for random initial bubble size}

\begin{figure}[htbp]
\centering
\begin{subfigure}{0.9\textwidth}
\includegraphics[width=1.0\linewidth]{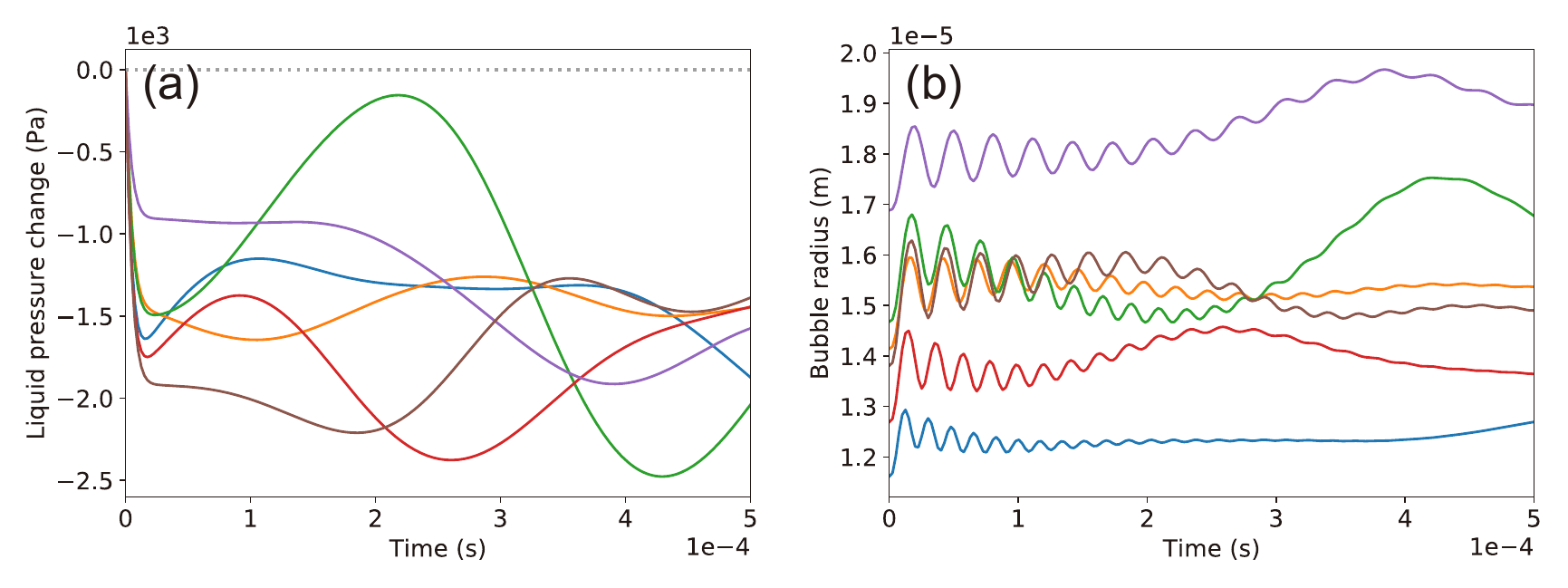}
\end{subfigure}
\caption{(a) Sample trajectories of liquid pressure change with time ($\Delta P(t)$); (b) Corresponding bubble radius ($R(t)$) with different initial size from the training dataset. The corresponding  $\Delta P(t)$ and $R(t)$ are presented with the same color. The initial bubble size has an effect on the oscillation frequency.}
\label{fig.exampleUS.randomR0}
\end{figure}

In this section, we extend the model to different initial bubble sizes.
The initial condition can be embedded in DeepONet in many ways, so here we choose to input $R(t_0)$ in 
the trunk net along with time $t$. 
In the current example, the training dataset contains 30,000 pairs of trajectories. Different from previous cases, the bubble radius trajectories are solved with random initial size ranging from $\SI{1e-5}{\meter}$ to $\SI{2e-5}{\meter}$. Some examples of the input and output functions are presented in Fig.~\ref{fig.exampleUS.randomR0}. As shown, the initial bubble size has a strong effect not only on the initial value of the solution but also on the oscillation frequency and the decay rate. 

\begin{figure}[h]
\begin{subfigure}{0.8\textwidth}
\includegraphics[width=1.0\linewidth]{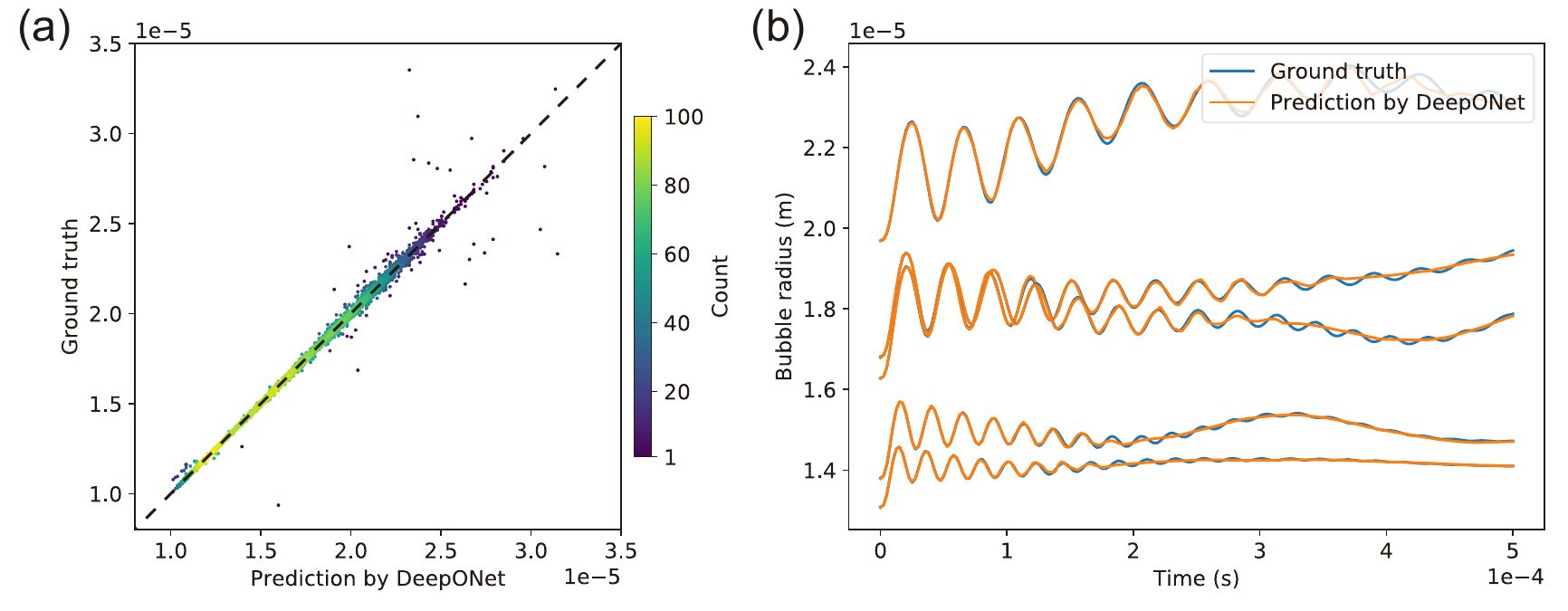}
\end{subfigure}
\caption{Comparison of DeepONet prediction and ground truth with random initial bubble size. (a) Parity plot of DeepONet prediction and ground truth from the test dataset; (b) Sample trajectories of prediction and ground truth from test dataset.}
\label{fig.randomR0.best}
\end{figure}

Upon training, DeepONet's predictions agree with the ground truth very well.
Fig.~\ref{fig.randomR0.best}(a) shows the comparison; each point's horizontal coordinate is a bubble radius value predicted by DeepONet at random times while the vertical coordinate is the corresponding ground truth. The color indicates the density of overlapping points. We see that most points lie on or near the diagonal line, which indicates an almost perfect prediction. 
Fig.~\ref{fig.randomR0.best}(b) shows some samples of bubble radius evolution predicted by DeepONet and solved with the R-P equation from the test dataset. We note that the different oscillation magnitude, oscillation frequency, and decay are accurately predicted by DeepONet.

\subsection{Extrapolation: Predicting inputs outside the function space of the training data}

The DeepONet predictions shown previously are results for pressure field in the input training range, i.e., the testing was performed with independent data but with pressure field generated with $l=0.2$ (Eq.~\eqref{eq.GRF}), the same
correlation length as in the training data (interpolation). In this section, we would like to obtain predictions with the trained DeepONet for pressure fields that do not belong to the input function space, i.e., perform an extrapolation. In particular, we want to use the trained DeepONet to predict the bubble radius based on pressure fields generated with unseen correlation length $l$,  ranging from $0.07$ to $0.9$. 
The other parameters used are as follows: $q$ in Eq.~\eqref{eq.GRF} equals $0.5$, the network size is $(50,50,50,50)$ for both trunk and branch net, and the training dataset contains 3000 pairs of trajectories. These settings remain the same through this section.  

\begin{figure}[ht]
\centering
\begin{subfigure}{0.9\textwidth}
\includegraphics[width=1.0\linewidth]{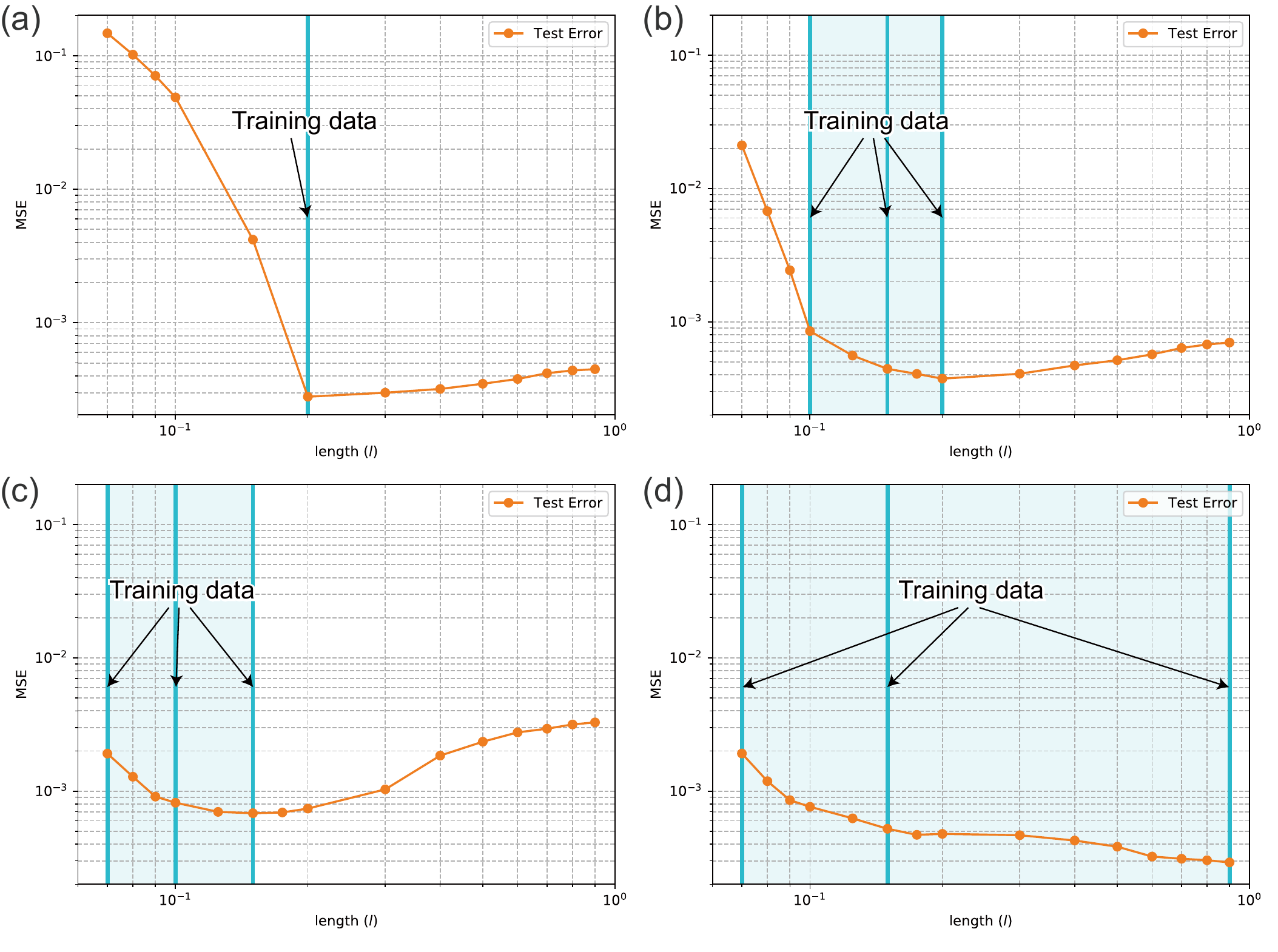}
\end{subfigure}
\caption{Extrapolation test error versus different choices of training datasets. (a) Training data: $l=0.2$; (b) Training data: $l=0.1,~0.15,~0.2$; (c) Training data: $l=0.07,~0.1,~0.15$; (d) Training data: $l=0.07,~0.15,~0.9$. The size of the training data remains the same in all cases. We note that by carefully designing the input space, the extrapolation performance can be significantly improved.}
\label{fig.extrapolation.error}
\end{figure}

In Fig.~\ref{fig.extrapolation.error}(a), we show the test error when we extrapolate  to unseen pressure field with $l$ ranging from $0.07$ to $0.9$. The error rises moderately when $l$ increases but may go up quite sharply when $l$ decreases.
In Fig.~\ref{fig.extrapolation.error}(b), we use as training data a combination of three different sets corresponding to values of $l$ ($0.1$, $0.15$ and $0.2$). Even though the training time and all other parameters are the same, the test errors of extrapolation are reduced substantially. The light blue area in the plot indicates the interpolation range; the test errors for $l=0.125$ and $l=0.175$ are smaller than that for $l=0.1$. 
Then, we design the representation of the input space more carefully, moving the window towards the smaller values of $l$, where we observed the highest error. 
In Fig.~\ref{fig.extrapolation.error}(c), the training data is a combination of $l=0.07$, $0.1$, and $0.15$. We observe that the errors with small $l$ are significantly reduced while for the extrapolation with large $l$ the errors increase. Overall, the errors in the entire range are more balanced. 
Fig.~\ref{fig.extrapolation.error}(d) shows an extra case when the training data is a combination of $l=0.07$, $0.15$, and $0.9$. We observe that the test errors are reduced more uniformly because the input space is better presented, which indicates that a better representation of the input space will lead to a higher accuracy for extrapolation.


\begin{figure}[bhtp]
\centering
\begin{subfigure}{0.95\textwidth}
\includegraphics[width=1.0\linewidth]{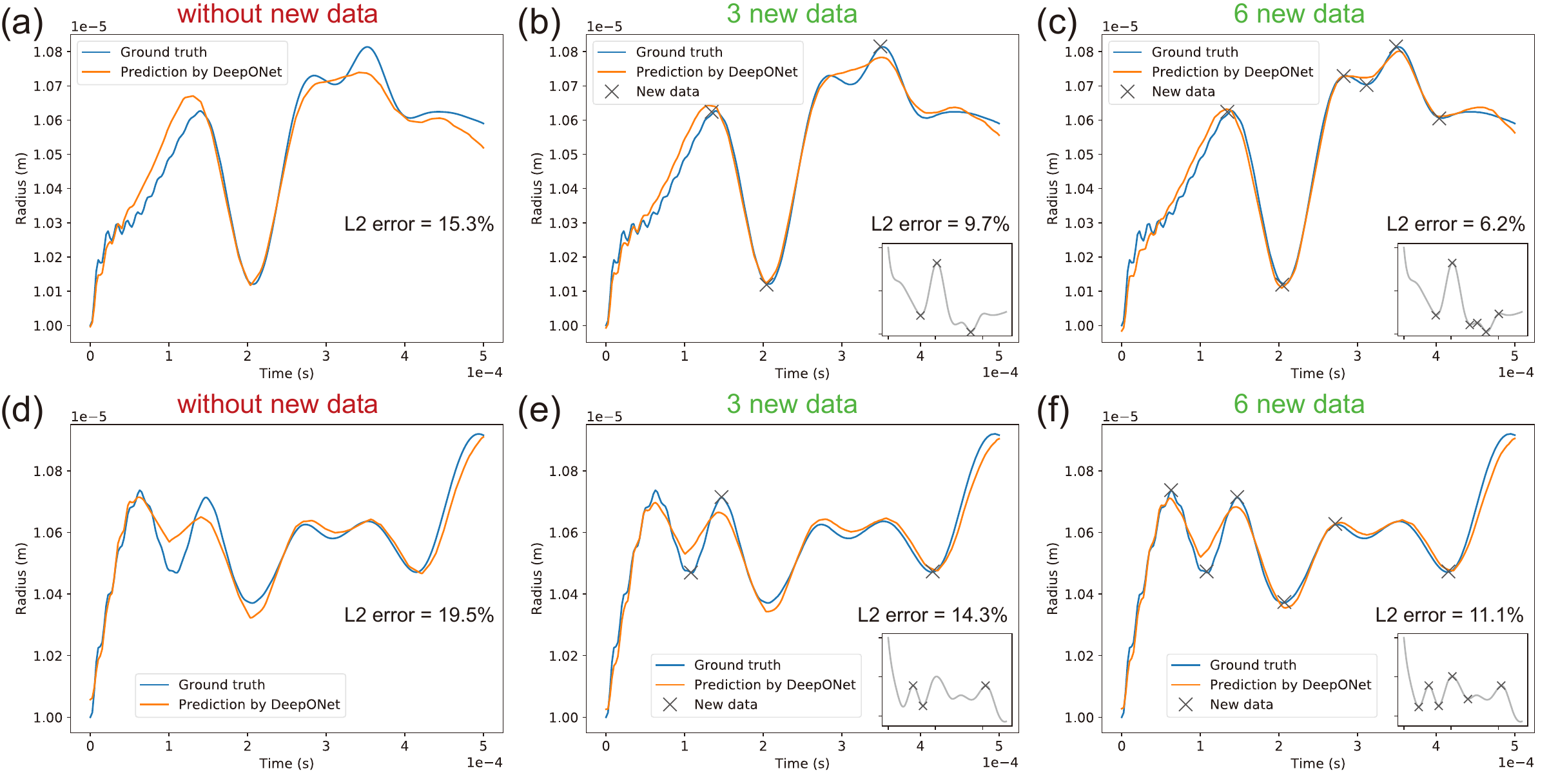}
\end{subfigure}
\caption{Extrapolations of $R(t)$ for $l=0.07$ by adding a few new data (Training data: $l=0.1,~0.15,~0.2$). (a-c) Example 1; (d-f) Example 2. For both examples, we show predictions without new data (a)(d), with 3 new data (b)(e), and with 6 new data (c)(f). The inset figures show the corresponding liquid pressure. The new data points are selected at the the pressure local extrema since the input function is known a priori. We 
fine-tune the network on these sparse new data points (black cross) with a reduced learning rate and epochs. Only the last layer of the trunk net is free to change by the new data. 
The $L_2$ error drops noticeably with just a few data for the extrapolation.}
\label{fig.extrapolation.l2}
\end{figure}

Next, we will demonstrate how extrapolation can be improved when only a few data are available for the output variable. In this example, the DeepONet is trained with liquid pressure produced with $l=0.1$, $0.15$, and $0.2$, as shown in Fig.~\ref{fig.extrapolation.error}(b), and is applied to extrapolate for $l=0.07$. We use the relative $L_2$-norm error to evaluate the performance, namely
\begin{equation}
{ \epsilon = \frac{\sqrt{\sum_{t=0}^{T}(R_{t}^{P}-R_{t}^{T})^2}} {\sqrt{\sum_{t=0}^{T}(R_{t}^{T}-\overline{R^{T}})^2}}}.
\end{equation}
where $R^P$ and $R^T$ are the predicted and true bubble radii, respectively. 
Fig.~\ref{fig.extrapolation.l2}(a)(c) show two examples of using the pre-trained DeepONet for making predictions on new scenarios (i.e., unseen data), where we can see that the $L_2$ errors are relatively high. 
In order to improve the performance of DeepONet for extrapolation, we assume that a few measurements of the unseen profile are available. Moreover, since the input function is known a priori, when doing extrapolation the locations of these additional data points can be selected based on the local extremum of the pressure profile. 
Then, we fine-tune the pre-trained networks based on only the new measurements by using the Adam optimizer with a reduced learning rate ($2\times10^{-5}$) for 100 epochs. We note that only the last layer of the trunk net is tunable during the fine-tuning process. The results are shown in Fig.~\ref{fig.extrapolation.l2}(b)(c)(e)(f), from which we can see that the predictions after fine-tuning are in good agreement with the ground truth.  
In the first case, the $L_2$ relative error drops from 15.3\% to 9.7\% (3 new data), or 6.2\% (6 new data), in the second case, the $L_2$ error drops from 19.5\% to 14.3\% (3 new data), or 11.1\% (6 new data). 
Overall, by performing transfer learning with sparse data points, we obtain much better solutions compared to just the original pretrained networks. In addition, we note that the transfer learning process is very fast, i.e., it takes only a few seconds either on a single CPU or GPU.

\section{Microscale (stochastic) regime}

The R-P equation is based on the  continuum assumption, which is not valid in the microscopic regime. Usually, particle-based simulation methods are good alternatives at the microscale. In this section, the DPD method is employed to simulate the dynamics of bubble size evolution at the microscale level and to generate data for training DeepONet.
As with MD systems, a DPD model consists of many interacting particles governed by Newton's equation of motion~\cite{groot1997dissipative},
%
\begin{equation}    \label{eq.DPD.EOM}
m_i\frac{{\rm d}^2\mathbf{r}_i}{{\rm d}t^2} = m_i\frac{{\rm d}\mathbf{v}_i}{{\rm d}t} = \mathbf{F}_i = \sum_{j\ne i}\left(\mathbf{F}^C_{ij} + \mathbf{F}^D_{ij} + \mathbf{F}^R_{ij}\right),
\end{equation}
%
where $m_i$ is the mass of a particle $i$, $\mathbf{r}_i$ and $\mathbf{v}_i$ are position and velocity vectors of the particle $i$, and $\mathbf{F}_i$ is the total force acting on the particle $i$ due to the presence of neighboring particles. The summation for computing $\mathbf{F}_i$ is carried out over all neighboring particles within a cutoff range. Beyond the cutoff, all pairwise interactions are considered negligible.
The pairwise force is comprised of a conservative force $\mathbf{F}^C_{ij}$, a dissipative force $\mathbf{F}^D_{ij}$ and a random force $\mathbf{F}^R_{ij}$, which, respectively, are of the form: \vspace{-0.2cm}

\begin{equation}    \label{eq.DPD.force}
\begin{split}
   & \mathbf{F}^C_{ij} = a_{ij}\omega_C(r_{ij})\mathbf{e}_{ij}, \\
   & \mathbf{F}^D_{ij} = -\gamma_{ij}\omega_D(r_{ij})(\mathbf{v}_{ij}\mathbf{e}_{ij})\mathbf{e}_{ij}, \\
   & \mathbf{F}^R_{ij}\cdot {\rm d}t = \sigma_{ij}\omega_R(r_{ij}){\rm d}\tilde{W}_{ij}\mathbf{e}_{ij},
\end{split}
\end{equation}
where $r_{ij}=|\mathbf{r}_{ij}| = |\mathbf{r}_i-\mathbf{r}_j|$ is the distance between particles $i$ and $j$, $\mathbf{e}_{ij}=\mathbf{r}_{ij}/r_{ij}$ is the unit vector, and $\mathbf{v}_{ij} = \mathbf{v}_i-\mathbf{v}_j$ is the velocity difference; ${\rm d}\tilde{W}_{ij}$ is an independent increment of the Wiener process~\cite{espanol1995statistical}. The coefficients $a_{ij}$, $\gamma_{ij}$ and $\sigma_{ij}$ determine the strength of the three forces, respectively. To satisfy the fluctuation-dissipation theorem (FDT)~\cite{espanol1995statistical} and to maintain the DPD system at a constant temperature~\cite{groot1997dissipative}, the dissipative force and the random force are constrained by $\sigma_{ij}^2 = 2\gamma_{ij}k_BT$ and $\omega_D(r_{ij})=\omega^2_R(r_{ij})$. Common choices for weight functions are $\omega_C(r_{ij}) = 1-r_{ij}/r_c$, $\omega_D(r_{ij}) = \omega^2_R(r_{ij}) = (1 - r_{ij}/r_c)^{1/2}$ with $r_c$ being the cutoff radius.

With the purely repulsive conservative forces, DPD is most suitable for simulating a gas, which fills the space spontaneously. To simulate the liquid phase, we use MDPD, which is an extension of DPD by modifying the conservative force to include both a repulsive force and an attractive force~\cite{warren2003vapor,arienti2011many},

\begin{equation}    \label{eq.MDPD.force}
\mathbf{F}_{ij}^C = A_{ij}w_c(r_{ij}) + B_{ij}(\rho_i + \rho_j) w_d(r_{ij}).
\end{equation}
Because a DPD model with a single interaction range cannot maintain a stable interface~\cite{pagonabarraga2001dissipative}, the repulsive contribution in Eq.~\eqref{eq.MDPD.force} is set to act within a shorter range $r_d < r_c$ than 
the soft pair attractive potential.
The many-body repulsion is chosen in the form of a self-energy per particle, which is quadratic in the local density, $B_{ij}(\rho_i+\rho_j)w_d(r_{ij})$, where $B>0$. The density of each particle is defined as

\begin{equation}    \label{eq.MDPD.rho}
\Bar{\rho_i} = \sum_{j \neq i}w_{\rho}(r_{ij}),
\end{equation}
and its weight function $w_{\rho}$ is defined as

\begin{equation}    \label{eq.MDPD.weight}
w_{\rho}=\frac{15}{2 \pi r_d^3}\left(1-\frac{r}{r_d}\right)^{2},
\end{equation}
where $w_\rho$ vanishes for $r > r_d$.

\begin{figure}[htb]
\centering
\begin{subfigure}{0.7\textwidth}
\includegraphics[width=1.0\linewidth]{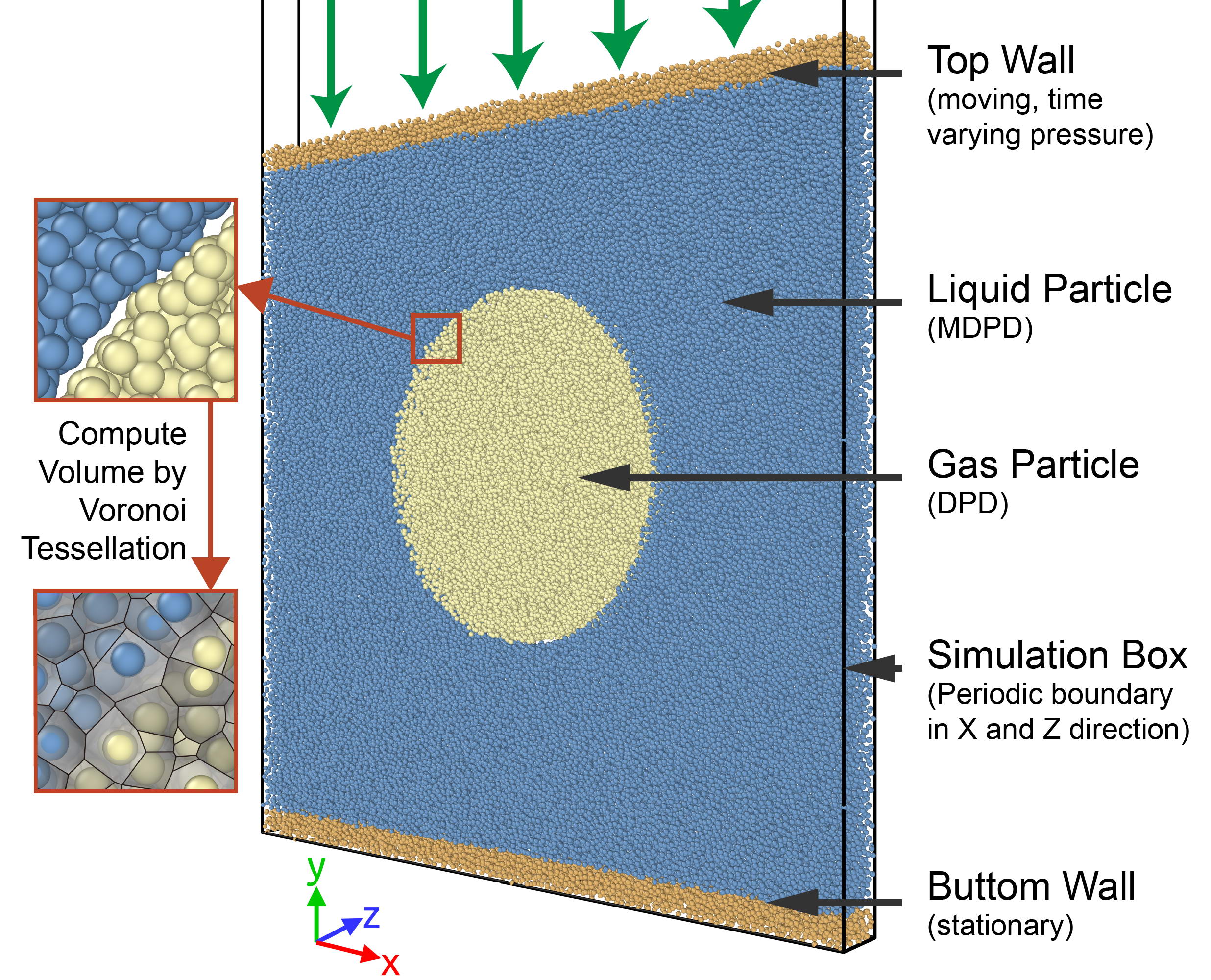}
\end{subfigure}
\caption{Illustration of the microscale simulation. The gas particles (yellow) are aggregated in the center surrounded by the liquid particles (blue). The boundary condition in the $x$ and $z$ directions are periodic. The bottom wall is stationary while the top wall has one degree of freedom in the $y$ direction to adjust the system pressure. A time-varying external force (green arrows) is applied on the top wall to change the system pressure. A Voronoi method is used to compute the volume of each phase.}
\label{fig.DPD}
\end{figure}

\begin{figure}[htbp]
\centering
\begin{subfigure}{0.4\textwidth}
\includegraphics[width=1.0\linewidth]{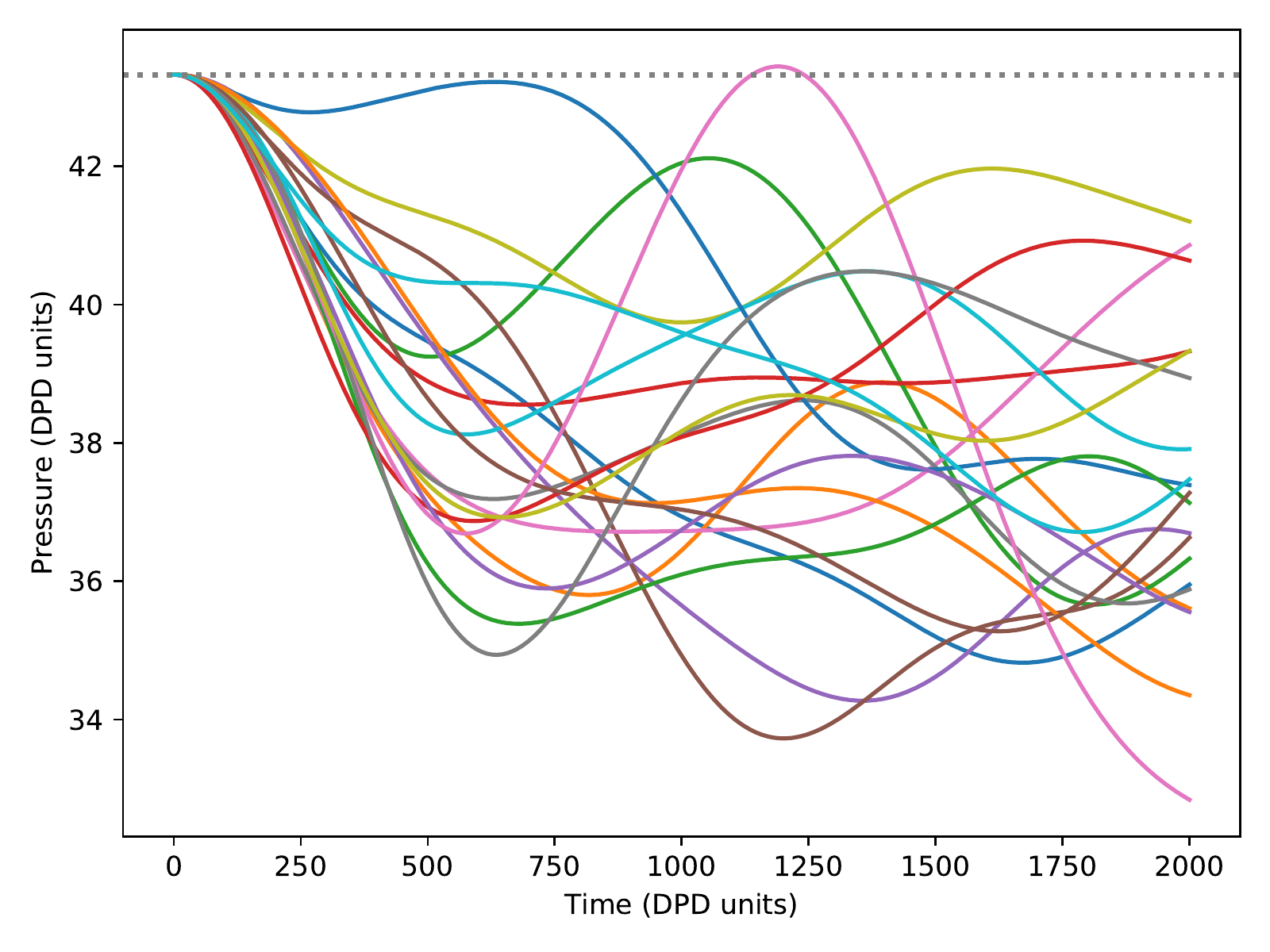}
\end{subfigure}
\caption{Samples of preset time-varying pressure on the top wall in training data. All trajectories start from an equilibrium pressure, then enter a stochastic time-varying low-pressure region. The dotted line indicates the initial equilibrium pressure.}
\label{fig.DPD.U}
\end{figure}

Fig.~\ref{fig.DPD} illustrates the DPD-MDPD coupled simulation system. To reduce the computational cost, a thin box with periodic boundaries in the $x$ and $z$ directions is used. The yellow particles represent the gas phase, with the interaction force modeled by a DPD potential; the blue particles represent the liquid phase, with the interaction force modeled by a MDPD potential. The interaction force between gas and liquid particles is described by a DPD potential, which can prevent the gas particles dissolving into the liquid phase. 
The top and bottom solid walls are constructed by pre-equilibrium frozen particles. 
The number density of wall particles is the same as the density of the liquid. 
The no-slip boundary method~\cite{li2018dissipative} for arbitrary-shape geometries is applied. To change the system pressure, the top wall is set to have one degree of freedom moving just in the $y$ direction. 
Before the simulation starts, a uniform static external pressure is exerted on the top wall to reach an initial equilibrium state. Then, the value of the external pressure evolves following a preset function. We continuously record the volume of the gas phase and use it to compute the effective bubble radius.
Because the particles of a DPD system are randomly distributed in space, it is difficult to accurately compute the volume of the gas bubble. In the present work, we adopt a three-dimensional Voronoi tessellation~\cite{rycroft2009voro} to estimate the instantaneous volume occupied by gas particles.
In the Voronoi scheme, each particle in the system occupies a cell around it, which contains all the space closer to that particle. In our case, the Voronoi cells take the form of irregular polyhedra, see the inset of Fig.~\ref{fig.DPD}.
Let $\Omega$ be the volume of a gas or liquid phase, we compute the instantaneous volume-averaged virial stress based on the virial theorem~\cite{tsai1979virial}, 
\begin{equation}
    S_{\alpha\beta}=\frac{1}{\Omega}\sum_{i\in\Omega}\left[-mv_{\alpha}^{i}v_{\beta}^{i} - \frac{1}{2}\sum^{N_p}_{j=1}(r^{i}_{\alpha}F^{i}_{\beta} + r^{j}_{\alpha}F^{j}_{\beta})\right],
\end{equation}
where $\alpha$ and $\beta$ take values of $x$, $y$, $z$ to generate the components of the stress. The first term represents a kinetic energy contribution for a particle $i$. The second term is a pairwise energy contribution due to interactions between particles $i$ and $j$, where $j$ loops over the $N_p$ neighbors of particle $i$; $\textbf{r}^{i}$ and $\textbf{r}^{j}$ are the positions of the two particles in the pairwise interaction, and $\textbf{F}^i$ and $\textbf{F}^j$ are the forces on the two particles resulting from the pairwise interaction.
We take the mean of diagonals of the negative stress tensor as the pressure:
\begin{equation}
    P_{phase} = -\frac{1}{3}(S_{xx} + S_{yy} + S_{zz}).
\end{equation}
Separate simulations are conducted to measure the properties of the gas and liquid phases. The simulation setup and the measured properties of fluids are summarised in Table~\ref{table.DPD.property}. 
By mapping the dimensionless properties (density, surface tension and viscosity) to properties of water at $\SI{20}{\celsius}$, the scaling relationship can be determined~\cite{lin2018tuning}: length reference $[l]=\SI{4.12e-8}{\meter}$, time reference $[t]=\SI{1.79e-9}{\second}$, and mass reference $[m]=\SI{8.82e-21}{\kilogram}$. 
With this scaling relationship, the dimensionless properties in the DPD simulation can be transformed to physical values. For example, the bubble radius is $\SI{5.77e-7}{\meter}$, and the simulation time span is approximately $\SI{3.58e-6}{\second}$.     
The preset pressure on the top wall is generated using GRF (Eq.~\eqref{eq.GRF}, $\mu=37.55$, $\sigma=2.17$) with an exponential function mask ensuring smooth transition from the initial equilibrium state. Examples of pressure trajectories are presented in Fig.~\ref{fig.DPD.U}.

\begin{figure}[htbp]
\centering
\begin{subfigure}{0.9\textwidth}
\includegraphics[width=1.0\linewidth]{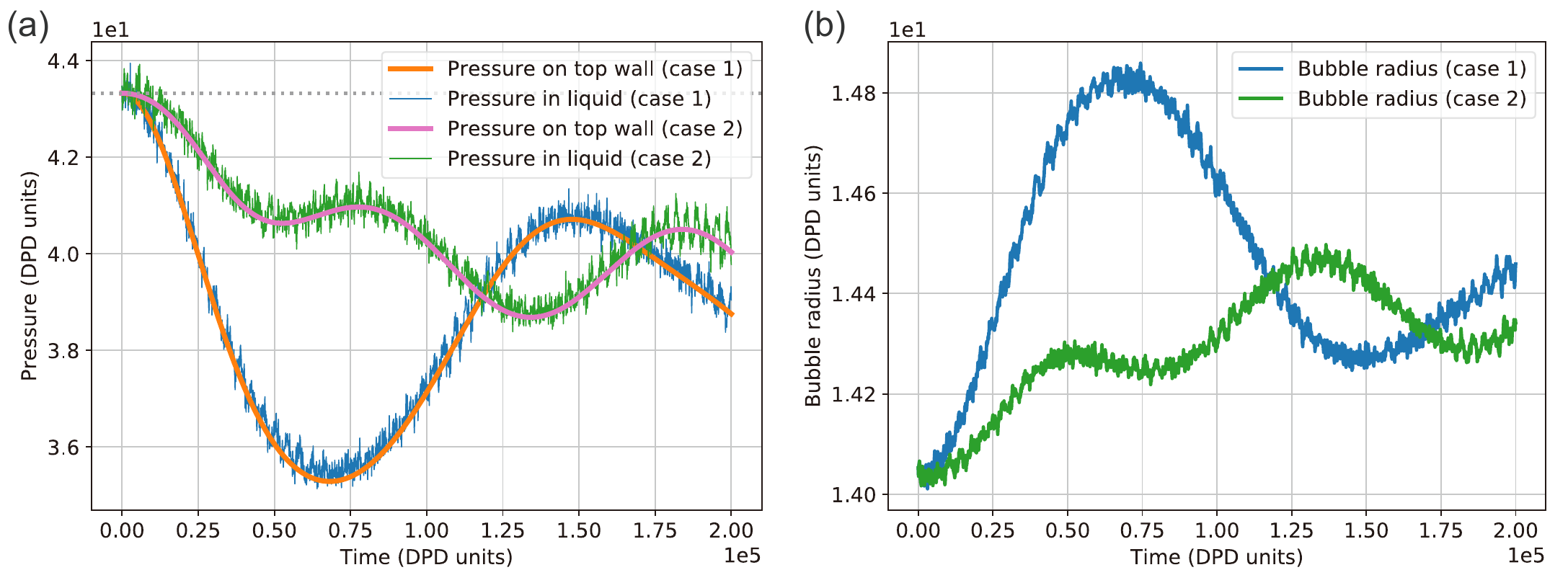}
\end{subfigure}
\caption{(a) Pressure on wall and pressure in liquid; (b) Corresponding bubble radius. The liquid pressure and bubble radius are both computed from DPD simulations and they exhibit considerable noise.}
\label{fig.DPD.1.U.s}
\end{figure}

We run 400 independent DPD simulations to generate the training data for DeepONet. Each simulation can generate one pair of input and output functions. Fig.~\ref{fig.DPD.1.U.s} shows two representative simulations. In Fig.~\ref{fig.DPD.1.U.s}(a), we plot both the preset pressure on the top wall and the measured liquid pressure. The measured liquid pressure exhibits noticeable noise compared to smooth input pressure signal. Fig.~\ref{fig.DPD.1.U.s}(b) shows the measured bubble radius; the corresponding $\Delta P(t)$ and $R(t)$ are presented with the same color.  

The setup of the DeepONet is summarized in Table~\ref{table.hyer-parameter}. Here, 50 evenly spaced sensors are employed for the input function and 25 random spaced points for the output function. Both branch and trunk nets have two hidden layers and 200 neurons per layer.

\clearpage
\begin{table}
\scriptsize
\centering
\caption{Simulation setup and fluid properties (DPD units).}
\begin{tabular}{lll} 
\hline
                           & Property           & Value                                         \\
\hline
\multirow{8}{*}{System}    & \multirow{3}{*}{Box} & $X=60$ (periodic)                           \\
                           &                      & $Y=66$ (shrink-wrapped)                     \\
                           &                      & $Z=4$ (periodic)                            \\
                           & Temperature          & $k_{b}T=1.0$                                \\
                           & Cutoff               & $r_{c}=1.0$                                 \\ 
                           & Time                 & $T=2000$                                    \\           
                           & Time step            & $t=0.01$                                    \\ 
                           & Pressure             & see Eq.~\eqref{eq.GRF}, $\mu=37.55$, $\sigma=2.17$     \\
\hline
\multirow{6}{*}{Liquid}    & MDPD parameter       & $A=-40$, $B=-25$, $\gamma=4.5$, $r_d=0.75$  \\
                           & Number density       & $n_{l}=7.93$                                \\
                           & Particle mass        & $m_{l}=1.0$                                 \\
                           & Liquid density       & $\rho_{l}=7.93$                             \\
                           & Dynamic viscosity    & $\mu_{l}=8.36$                              \\
                           & Speed of sound       & $c=18.52$                                   \\ 
\hline
\multirow{5}{*}{Gas}       & DPD parameter      & $a=5$,  $\gamma=4.5$                          \\
                           & Number density     & $n_{g}=8.25$                                  \\
                           & Particle mass      & $m_{g}=0.2$                                   \\
                           & Gas density        & $\rho_{g}=1.65$                               \\
                           & Dynamic viscosity  & $\mu_{g}=1.41$                                \\ 
\hline
\multirow{5}{*}{Interface} & DPD parameter      & $a=20$,  $\gamma=4.5$                         \\
                           & Pressure in liquid & $p_{l}=43.08$                                 \\
                           & Pressure in gas    & $p_{g}=44.96$                                 \\
                           & Bubble radius      & $R=14.05$                                     \\
                           & Surface tension    & $\gamma=26.42$                                \\
\hline
                           &                    &                                                 
\end{tabular}
\label{table.DPD.property}
\end{table}

\begin{table}
\scriptsize
\centering
\caption{Hyper-parameters for DeepONet.}
\label{table.hyer-parameter}
\begin{tabular}{|p{3.5cm}|p{3.2cm}|p{2.75cm}|p{2.5cm}|p{2.4cm}|} 
\hline
\# Pressure (Train+Test)        & \# Pressure sensor  & \# Radius sensor & \# Train data & \# Test data  \\
400+100                   & 50                         & 25               & 10000         & 2500          \\ 
\hline
Type of DeepONet          & Branch depth               & Branch width     & Trunk depth   & Trunk width   \\
Unstacked                 & 2                          & 200              & 2             & 200           \\ 
\hline
Pressure generator & Optimizer & Learning rate       & Activation & \# Epoch      \\
Gaussian random           & Adam           & 0.0005             & ReLU        & 10000         \\
\hline
\end{tabular}
\end{table}
\clearpage

\begin{figure}[htbp]
\centering
\begin{subfigure}{0.9\textwidth}
\includegraphics[width=1.0\linewidth]{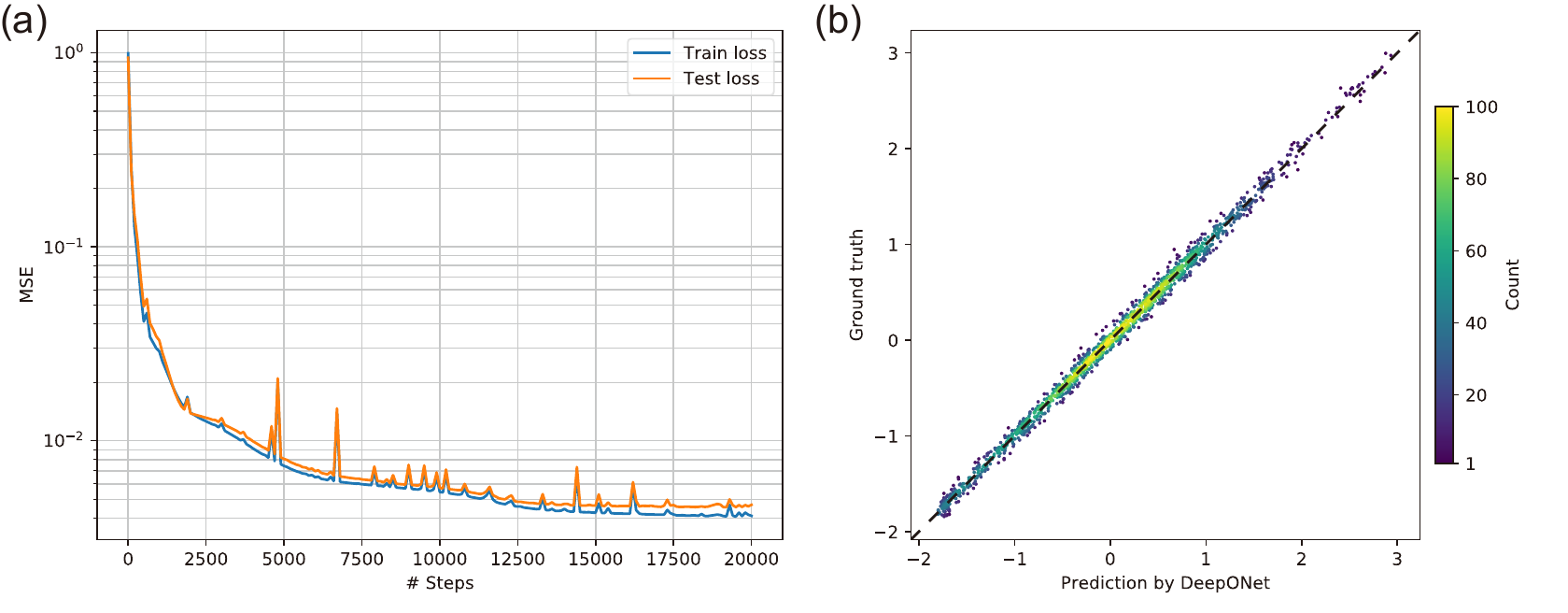}
\end{subfigure}
\caption{(a) Loss histories of the training and testing; (b) Parity plot of DeepONet prediction and DPD simulation results; the data points fall on the diagonal dashed line indicating very good prediction.}
\label{fig.DPD.loss}
\end{figure}

\begin{figure}[htbp]
\centering
\begin{subfigure}{0.9\textwidth}
\includegraphics[width=1.0\linewidth]{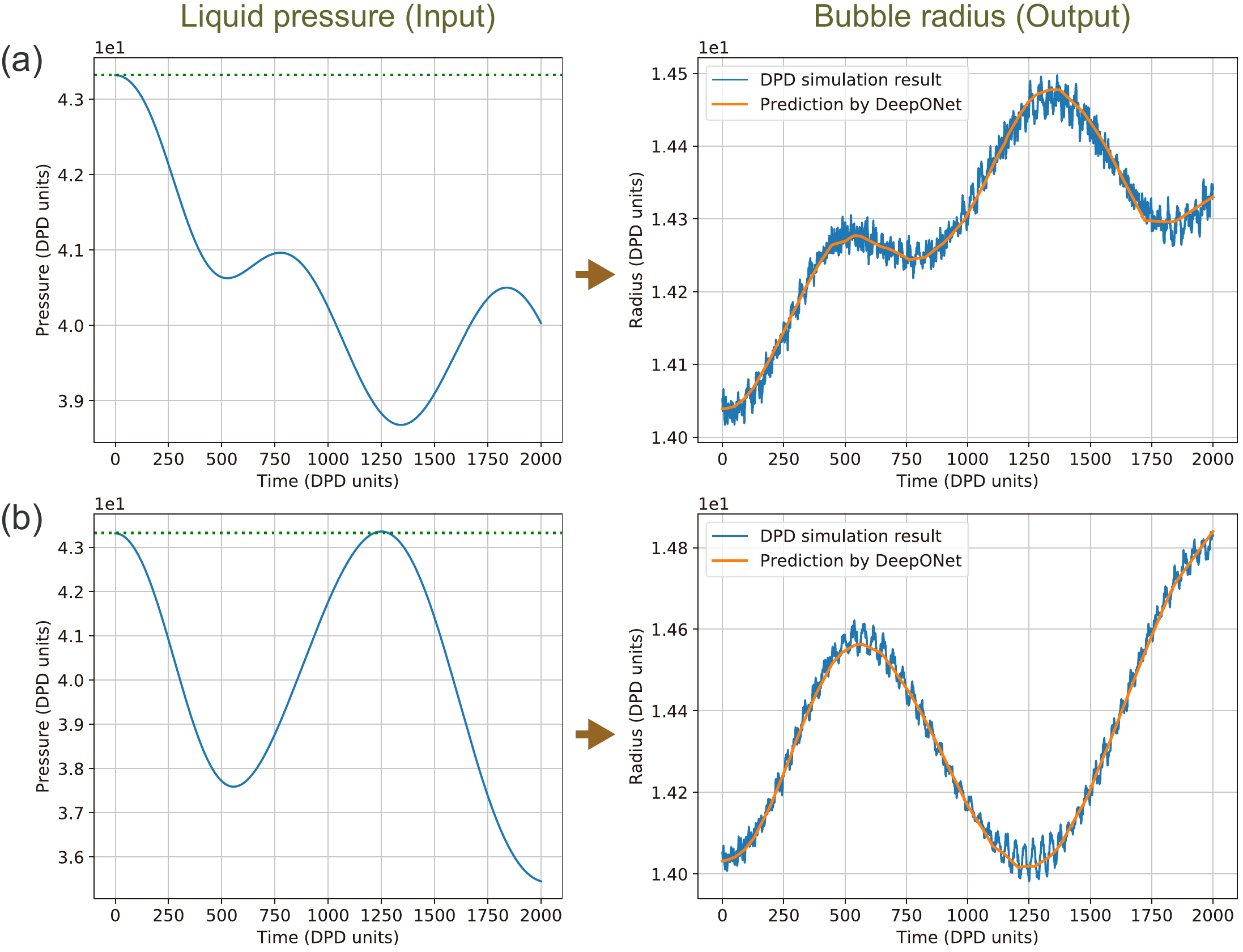}
\end{subfigure}
\caption{Two representative DeepONet predictions (a and b). The plots in the left column correspond to the preset pressure; the plots in the right column show a comparison of the bubble radius produced by DPD simulations and predicted by DeepONet. DeepONet's prediction is close to the mean of the fluctuating trajectory.}
\label{fig.DPD.compare}
\end{figure}

Fig.~\ref{fig.DPD.loss}(a) shows the loss histories for training and testing. 
Fig.~\ref{fig.DPD.loss}(b) is the parity plot of values predicted by DeepONet and DPD simulation results. We see that almost all data points fall near the diagonal line, which indicates almost perfect accuracy. The small discrepancy is mainly because the raw trajectory is noisy but the predicted trajectory is quite smooth, see Fig.~\ref{fig.DPD.compare}.

Fig.~\ref{fig.DPD.compare} shows two representative cases of the DeepONet prediction. Even though the DPD results exhibit high frequency fluctuations and we only collect sparse data points on the noisy trajectory, DeepONet yields an accurate smooth trajectory, which captures the mean trajectory of the data quite well.

The results shown above indicate that once trained, DeepONet can be used to infer the microscale bubble growth from a complex particle simulation system, which cannot be easily described by simple control functions. Conventional numerical simulation methods, in which pair-list updates and inter-particle force calculation are required, are usually very time consuming. For each DPD simulation, it takes 3 hours on a 16-core workstation, while it takes only a fraction of a second for DeepONet to evaluate $R(t)$, so the  speed-up factor in this case exceeds 200,000.   

\section{Summary}
In this paper, we first introduce the macroscopic model (R-P) and microscopic model (DPD) for the problem of growth of a  single bubble due to time-varying changes in the ambient liquid pressure. We then demonstrate the effectiveness of DeepONet for both models. 
We use Gaussian random fields to generate the various pressure fields, which are the input signals to the dynamical system. We investigate the influence of the network depth, width, and data size, and by comparing to LSTM we demonstrated a unique advantage of DeepONet, i.e.,  the flexibility in data density and location, which will be very useful when we have no prior knowledge of how much data density is sufficient or when we have restrictions on data acquisition. 
Moreover, we test the case when the input is not within the input space, i.e., the correlation length of the pressure field is out of the training range. In this case, we cannot obtain very accurate predictions, however, we resolve this issue by  transfer learning and fine-tuning the trunk net of the pre-trained DeepONet with just a few new data points. 
In the microscale regime, the data generated by DPD model contain high frequency fluctuations. We show that without any extra data processing, DeepONet can learn the mean component of these noisy raw data, and the computational time could be reduced from 48 (3 hours $\times$ 16 cores) CPU hours to a fraction of second for a speed-up over 200,000. Learning the stochastic fluctuations accurately requires a more careful design and training of DeepONet and we leave this for future work

More broadly, the results in this paper show that DeepONet can be used both in the macroscopic and microscopic regimes of the bubble growth dynamics, which creates a good foundation for a unified neural network model that could predict the results from macroscale to microscale seamlessly. This topic is currently under investigation and we will report such results in future work.

\section*{Acknowledgements} %
The authors acknowledge support from DOE/PhILMs DE-SC0019453.

\section*{DATA AVAILABILITY} %
The data that support the findings of this study are available from the corresponding author upon reasonable request.

\bibliography{ref.bib}

\begin{thebibliography}{47}%
\makeatletter
\providecommand \@ifxundefined [1]{%
 \@ifx{#1\undefined}
}%
\providecommand \@ifnum [1]{%
 \ifnum #1\expandafter \@firstoftwo
 \else \expandafter \@secondoftwo
 \fi
}%
\providecommand \@ifx [1]{%
 \ifx #1\expandafter \@firstoftwo
 \else \expandafter \@secondoftwo
 \fi
}%
\providecommand \natexlab [1]{#1}%
\providecommand \enquote  [1]{``#1''}%
\providecommand \bibnamefont  [1]{#1}%
\providecommand \bibfnamefont [1]{#1}%
\providecommand \citenamefont [1]{#1}%
\providecommand \href@noop [0]{\@secondoftwo}%
\providecommand \href [0]{\begingroup \@sanitize@url \@href}%
\providecommand \@href[1]{\@@startlink{#1}\@@href}%
\providecommand \@@href[1]{\endgroup#1\@@endlink}%
\providecommand \@sanitize@url [0]{\catcode `\\12\catcode `\$12\catcode
  `\&12\catcode `\#12\catcode `\^12\catcode `\_12\catcode `\%12\relax}%
\providecommand \@@startlink[1]{}%
\providecommand \@@endlink[0]{}%
\providecommand \url  [0]{\begingroup\@sanitize@url \@url }%
\providecommand \@url [1]{\endgroup\@href {#1}{\urlprefix }}%
\providecommand \urlprefix  [0]{URL }%
\providecommand \Eprint [0]{\href }%
\providecommand \doibase [0]{http://dx.doi.org/}%
\providecommand \selectlanguage [0]{\@gobble}%
\providecommand \bibinfo  [0]{\@secondoftwo}%
\providecommand \bibfield  [0]{\@secondoftwo}%
\providecommand \translation [1]{[#1]}%
\providecommand \BibitemOpen [0]{}%
\providecommand \bibitemStop [0]{}%
\providecommand \bibitemNoStop [0]{.\EOS\space}%
\providecommand \EOS [0]{\spacefactor3000\relax}%
\providecommand \BibitemShut  [1]{\csname bibitem#1\endcsname}%
\let\auto@bib@innerbib\@empty
\bibitem [{\citenamefont {Rudy}\ \emph {et~al.}(2017)\citenamefont {Rudy},
  \citenamefont {Brunton}, \citenamefont {Proctor},\ and\ \citenamefont
  {Kutz}}]{rudy2017data}%
  \BibitemOpen
  \bibfield  {author} {\bibinfo {author} {\bibfnamefont {S.~H.}\ \bibnamefont
  {Rudy}}, \bibinfo {author} {\bibfnamefont {S.~L.}\ \bibnamefont {Brunton}},
  \bibinfo {author} {\bibfnamefont {J.~L.}\ \bibnamefont {Proctor}}, \ and\
  \bibinfo {author} {\bibfnamefont {J.~N.}\ \bibnamefont {Kutz}},\ }\bibfield
  {title} {\enquote {\bibinfo {title} {Data-driven discovery of partial
  differential equations},}\ }\href@noop {} {\bibfield  {journal} {\bibinfo
  {journal} {Science Advances}\ }\textbf {\bibinfo {volume} {3}},\ \bibinfo
  {pages} {e1602614} (\bibinfo {year} {2017})}\BibitemShut {NoStop}%
\bibitem [{\citenamefont {Champion}\ \emph {et~al.}(2019)\citenamefont
  {Champion}, \citenamefont {Lusch}, \citenamefont {Kutz},\ and\ \citenamefont
  {Brunton}}]{champion2019data}%
  \BibitemOpen
  \bibfield  {author} {\bibinfo {author} {\bibfnamefont {K.}~\bibnamefont
  {Champion}}, \bibinfo {author} {\bibfnamefont {B.}~\bibnamefont {Lusch}},
  \bibinfo {author} {\bibfnamefont {J.~N.}\ \bibnamefont {Kutz}}, \ and\
  \bibinfo {author} {\bibfnamefont {S.~L.}\ \bibnamefont {Brunton}},\
  }\bibfield  {title} {\enquote {\bibinfo {title} {Data-driven discovery of
  coordinates and governing equations},}\ }\href@noop {} {\bibfield  {journal}
  {\bibinfo  {journal} {Proceedings of the National Academy of Sciences}\
  }\textbf {\bibinfo {volume} {116}},\ \bibinfo {pages} {22445--22451}
  (\bibinfo {year} {2019})}\BibitemShut {NoStop}%
\bibitem [{\citenamefont {Lu}, \citenamefont {Jin},\ and\ \citenamefont
  {Karniadakis}(2019)}]{lu2019deeponet}%
  \BibitemOpen
  \bibfield  {author} {\bibinfo {author} {\bibfnamefont {L.}~\bibnamefont
  {Lu}}, \bibinfo {author} {\bibfnamefont {P.}~\bibnamefont {Jin}}, \ and\
  \bibinfo {author} {\bibfnamefont {G.~E.}\ \bibnamefont {Karniadakis}},\
  }\bibfield  {title} {\enquote {\bibinfo {title} {{DeepONet}: Learning
  nonlinear operators for identifying differential equations based on the
  universal approximation theorem of operators},}\ }\href@noop {} {\bibfield
  {journal} {\bibinfo  {journal} {arXiv preprint arXiv:1910.03193}\ } (\bibinfo
  {year} {2019})}\BibitemShut {NoStop}%
\bibitem [{\citenamefont {Raissi}, \citenamefont {Yazdani},\ and\ \citenamefont
  {Karniadakis}(2020)}]{raissi2020hidden}%
  \BibitemOpen
  \bibfield  {author} {\bibinfo {author} {\bibfnamefont {M.}~\bibnamefont
  {Raissi}}, \bibinfo {author} {\bibfnamefont {A.}~\bibnamefont {Yazdani}}, \
  and\ \bibinfo {author} {\bibfnamefont {G.~E.}\ \bibnamefont {Karniadakis}},\
  }\bibfield  {title} {\enquote {\bibinfo {title} {Hidden fluid mechanics:
  Learning velocity and pressure fields from flow visualizations},}\
  }\href@noop {} {\bibfield  {journal} {\bibinfo  {journal} {Science}\ }\textbf
  {\bibinfo {volume} {367}},\ \bibinfo {pages} {1026--1030} (\bibinfo {year}
  {2020})}\BibitemShut {NoStop}%
\bibitem [{\citenamefont {Raissi}, \citenamefont {Perdikaris},\ and\
  \citenamefont {Karniadakis}(2019)}]{raissi2019physics}%
  \BibitemOpen
  \bibfield  {author} {\bibinfo {author} {\bibfnamefont {M.}~\bibnamefont
  {Raissi}}, \bibinfo {author} {\bibfnamefont {P.}~\bibnamefont {Perdikaris}},
  \ and\ \bibinfo {author} {\bibfnamefont {G.~E.}\ \bibnamefont
  {Karniadakis}},\ }\bibfield  {title} {\enquote {\bibinfo {title}
  {Physics-informed neural networks: A deep learning framework for solving
  forward and inverse problems involving nonlinear partial differential
  equations},}\ }\href@noop {} {\bibfield  {journal} {\bibinfo  {journal}
  {Journal of Computational Physics}\ }\textbf {\bibinfo {volume} {378}},\
  \bibinfo {pages} {686--707} (\bibinfo {year} {2019})}\BibitemShut {NoStop}%
\bibitem [{\citenamefont {Jin}\ \emph {et~al.}(2020)\citenamefont {Jin},
  \citenamefont {Zhang}, \citenamefont {Zhu}, \citenamefont {Tang},\ and\
  \citenamefont {Karniadakis}}]{jin2020sympnets}%
  \BibitemOpen
  \bibfield  {author} {\bibinfo {author} {\bibfnamefont {P.}~\bibnamefont
  {Jin}}, \bibinfo {author} {\bibfnamefont {Z.}~\bibnamefont {Zhang}}, \bibinfo
  {author} {\bibfnamefont {A.}~\bibnamefont {Zhu}}, \bibinfo {author}
  {\bibfnamefont {Y.}~\bibnamefont {Tang}}, \ and\ \bibinfo {author}
  {\bibfnamefont {G.~E.}\ \bibnamefont {Karniadakis}},\ }\bibfield  {title}
  {\enquote {\bibinfo {title} {Sympnets: Intrinsic structure-preserving
  symplectic networks for identifying hamiltonian systems},}\ }\href@noop {}
  {\bibfield  {journal} {\bibinfo  {journal} {Neural Networks}\ }\textbf
  {\bibinfo {volume} {132}},\ \bibinfo {pages} {166--179} (\bibinfo {year}
  {2020})}\BibitemShut {NoStop}%
\bibitem [{\citenamefont {Chen}\ \emph {et~al.}(2020)\citenamefont {Chen},
  \citenamefont {Lu}, \citenamefont {Karniadakis},\ and\ \citenamefont
  {Dal~Negro}}]{chen2020physics}%
  \BibitemOpen
  \bibfield  {author} {\bibinfo {author} {\bibfnamefont {Y.}~\bibnamefont
  {Chen}}, \bibinfo {author} {\bibfnamefont {L.}~\bibnamefont {Lu}}, \bibinfo
  {author} {\bibfnamefont {G.~E.}\ \bibnamefont {Karniadakis}}, \ and\ \bibinfo
  {author} {\bibfnamefont {L.}~\bibnamefont {Dal~Negro}},\ }\bibfield  {title}
  {\enquote {\bibinfo {title} {Physics-informed neural networks for inverse
  problems in nano-optics and metamaterials},}\ }\href@noop {} {\bibfield
  {journal} {\bibinfo  {journal} {Optics Express}\ }\textbf {\bibinfo {volume}
  {28}},\ \bibinfo {pages} {11618--11633} (\bibinfo {year} {2020})}\BibitemShut
  {NoStop}%
\bibitem [{\citenamefont {Carleo}\ \emph {et~al.}(2019)\citenamefont {Carleo},
  \citenamefont {Cirac}, \citenamefont {Cranmer}, \citenamefont {Daudet},
  \citenamefont {Schuld}, \citenamefont {Tishby}, \citenamefont
  {Vogt-Maranto},\ and\ \citenamefont {Zdeborov{\'a}}}]{carleo2019machine}%
  \BibitemOpen
  \bibfield  {author} {\bibinfo {author} {\bibfnamefont {G.}~\bibnamefont
  {Carleo}}, \bibinfo {author} {\bibfnamefont {I.}~\bibnamefont {Cirac}},
  \bibinfo {author} {\bibfnamefont {K.}~\bibnamefont {Cranmer}}, \bibinfo
  {author} {\bibfnamefont {L.}~\bibnamefont {Daudet}}, \bibinfo {author}
  {\bibfnamefont {M.}~\bibnamefont {Schuld}}, \bibinfo {author} {\bibfnamefont
  {N.}~\bibnamefont {Tishby}}, \bibinfo {author} {\bibfnamefont
  {L.}~\bibnamefont {Vogt-Maranto}}, \ and\ \bibinfo {author} {\bibfnamefont
  {L.}~\bibnamefont {Zdeborov{\'a}}},\ }\bibfield  {title} {\enquote {\bibinfo
  {title} {Machine learning and the physical sciences},}\ }\href@noop {}
  {\bibfield  {journal} {\bibinfo  {journal} {Reviews of Modern Physics}\
  }\textbf {\bibinfo {volume} {91}},\ \bibinfo {pages} {045002} (\bibinfo
  {year} {2019})}\BibitemShut {NoStop}%
\bibitem [{\citenamefont {Cichy}\ and\ \citenamefont
  {Kaiser}(2019)}]{cichy2019deep}%
  \BibitemOpen
  \bibfield  {author} {\bibinfo {author} {\bibfnamefont {R.~M.}\ \bibnamefont
  {Cichy}}\ and\ \bibinfo {author} {\bibfnamefont {D.}~\bibnamefont {Kaiser}},\
  }\bibfield  {title} {\enquote {\bibinfo {title} {Deep neural networks as
  scientific models},}\ }\href@noop {} {\bibfield  {journal} {\bibinfo
  {journal} {Trends in Cognitive Sciences}\ }\textbf {\bibinfo {volume} {23}},\
  \bibinfo {pages} {305--317} (\bibinfo {year} {2019})}\BibitemShut {NoStop}%
\bibitem [{\citenamefont {Lapeyre}\ \emph {et~al.}(2019)\citenamefont
  {Lapeyre}, \citenamefont {Misdariis}, \citenamefont {Cazard}, \citenamefont
  {Veynante},\ and\ \citenamefont {Poinsot}}]{lapeyre2019training}%
  \BibitemOpen
  \bibfield  {author} {\bibinfo {author} {\bibfnamefont {C.~J.}\ \bibnamefont
  {Lapeyre}}, \bibinfo {author} {\bibfnamefont {A.}~\bibnamefont {Misdariis}},
  \bibinfo {author} {\bibfnamefont {N.}~\bibnamefont {Cazard}}, \bibinfo
  {author} {\bibfnamefont {D.}~\bibnamefont {Veynante}}, \ and\ \bibinfo
  {author} {\bibfnamefont {T.}~\bibnamefont {Poinsot}},\ }\bibfield  {title}
  {\enquote {\bibinfo {title} {Training convolutional neural networks to
  estimate turbulent sub-grid scale reaction rates},}\ }\href@noop {}
  {\bibfield  {journal} {\bibinfo  {journal} {Combustion and Flame}\ }\textbf
  {\bibinfo {volume} {203}},\ \bibinfo {pages} {255--264} (\bibinfo {year}
  {2019})}\BibitemShut {NoStop}%
\bibitem [{\citenamefont {Bieker}\ \emph {et~al.}(2020)\citenamefont {Bieker},
  \citenamefont {Peitz}, \citenamefont {Brunton}, \citenamefont {Kutz},\ and\
  \citenamefont {Dellnitz}}]{bieker2020deep}%
  \BibitemOpen
  \bibfield  {author} {\bibinfo {author} {\bibfnamefont {K.}~\bibnamefont
  {Bieker}}, \bibinfo {author} {\bibfnamefont {S.}~\bibnamefont {Peitz}},
  \bibinfo {author} {\bibfnamefont {S.~L.}\ \bibnamefont {Brunton}}, \bibinfo
  {author} {\bibfnamefont {J.~N.}\ \bibnamefont {Kutz}}, \ and\ \bibinfo
  {author} {\bibfnamefont {M.}~\bibnamefont {Dellnitz}},\ }\bibfield  {title}
  {\enquote {\bibinfo {title} {Deep model predictive flow control with limited
  sensor data and online learning},}\ }\href@noop {} {\bibfield  {journal}
  {\bibinfo  {journal} {Theoretical and Computational Fluid Dynamics}\ ,\
  \bibinfo {pages} {1--15}} (\bibinfo {year} {2020})}\BibitemShut {NoStop}%
\bibitem [{\citenamefont {Xie}\ \emph {et~al.}(2018)\citenamefont {Xie},
  \citenamefont {Franz}, \citenamefont {Chu},\ and\ \citenamefont
  {Thuerey}}]{xie2018TempoGan}%
  \BibitemOpen
  \bibfield  {author} {\bibinfo {author} {\bibfnamefont {Y.}~\bibnamefont
  {Xie}}, \bibinfo {author} {\bibfnamefont {E.}~\bibnamefont {Franz}}, \bibinfo
  {author} {\bibfnamefont {M.}~\bibnamefont {Chu}}, \ and\ \bibinfo {author}
  {\bibfnamefont {N.}~\bibnamefont {Thuerey}},\ }\bibfield  {title} {\enquote
  {\bibinfo {title} {{tempoGAN}: A temporally coherent, volumetric gan for
  super-resolution fluid flow},}\ }\href@noop {} {\bibfield  {journal}
  {\bibinfo  {journal} {ACM Transactions on Graphics (TOG)}\ }\textbf {\bibinfo
  {volume} {37}},\ \bibinfo {pages} {1--15} (\bibinfo {year}
  {2018})}\BibitemShut {NoStop}%
\bibitem [{\citenamefont {Peng}\ \emph {et~al.}(2020)\citenamefont {Peng},
  \citenamefont {Alber}, \citenamefont {Tepole}, \citenamefont {Cannon},
  \citenamefont {De}, \citenamefont {Dura-Bernal}, \citenamefont {Garikipati},
  \citenamefont {Karniadakis}, \citenamefont {Lytton}, \citenamefont
  {Perdikaris} \emph {et~al.}}]{peng2020multiscale}%
  \BibitemOpen
  \bibfield  {author} {\bibinfo {author} {\bibfnamefont {G.~C.}\ \bibnamefont
  {Peng}}, \bibinfo {author} {\bibfnamefont {M.}~\bibnamefont {Alber}},
  \bibinfo {author} {\bibfnamefont {A.~B.}\ \bibnamefont {Tepole}}, \bibinfo
  {author} {\bibfnamefont {W.~R.}\ \bibnamefont {Cannon}}, \bibinfo {author}
  {\bibfnamefont {S.}~\bibnamefont {De}}, \bibinfo {author} {\bibfnamefont
  {S.}~\bibnamefont {Dura-Bernal}}, \bibinfo {author} {\bibfnamefont
  {K.}~\bibnamefont {Garikipati}}, \bibinfo {author} {\bibfnamefont {G.~E.}\
  \bibnamefont {Karniadakis}}, \bibinfo {author} {\bibfnamefont {W.~W.}\
  \bibnamefont {Lytton}}, \bibinfo {author} {\bibfnamefont {P.}~\bibnamefont
  {Perdikaris}},  \emph {et~al.},\ }\bibfield  {title} {\enquote {\bibinfo
  {title} {Multiscale modeling meets machine learning: What can we learn?}}\
  }\href@noop {} {\bibfield  {journal} {\bibinfo  {journal} {Archives of
  Computational Methods in Engineering}\ ,\ \bibinfo {pages} {1--21}} (\bibinfo
  {year} {2020})}\BibitemShut {NoStop}%
\bibitem [{\citenamefont {Mao}, \citenamefont {Jagtap},\ and\ \citenamefont
  {Karniadakis}(2020)}]{mao2020physics}%
  \BibitemOpen
  \bibfield  {author} {\bibinfo {author} {\bibfnamefont {Z.}~\bibnamefont
  {Mao}}, \bibinfo {author} {\bibfnamefont {A.~D.}\ \bibnamefont {Jagtap}}, \
  and\ \bibinfo {author} {\bibfnamefont {G.~E.}\ \bibnamefont {Karniadakis}},\
  }\bibfield  {title} {\enquote {\bibinfo {title} {Physics-informed neural
  networks for high-speed flows},}\ }\href@noop {} {\bibfield  {journal}
  {\bibinfo  {journal} {Computer Methods in Applied Mechanics and Engineering}\
  }\textbf {\bibinfo {volume} {360}},\ \bibinfo {pages} {112789} (\bibinfo
  {year} {2020})}\BibitemShut {NoStop}%
\bibitem [{\citenamefont {Yu}\ \emph {et~al.}(2017)\citenamefont {Yu},
  \citenamefont {Zhang}, \citenamefont {Wang},\ and\ \citenamefont
  {Jiang}}]{yu2017superwettability}%
  \BibitemOpen
  \bibfield  {author} {\bibinfo {author} {\bibfnamefont {C.}~\bibnamefont
  {Yu}}, \bibinfo {author} {\bibfnamefont {P.}~\bibnamefont {Zhang}}, \bibinfo
  {author} {\bibfnamefont {J.}~\bibnamefont {Wang}}, \ and\ \bibinfo {author}
  {\bibfnamefont {L.}~\bibnamefont {Jiang}},\ }\bibfield  {title} {\enquote
  {\bibinfo {title} {Superwettability of gas bubbles and its application: from
  bioinspiration to advanced materials},}\ }\href@noop {} {\bibfield  {journal}
  {\bibinfo  {journal} {Advanced Materials}\ }\textbf {\bibinfo {volume}
  {29}},\ \bibinfo {pages} {1703053} (\bibinfo {year} {2017})}\BibitemShut
  {NoStop}%
\bibitem [{\citenamefont {Dollet}, \citenamefont {Marmottant},\ and\
  \citenamefont {Garbin}(2019)}]{dollet2019bubble}%
  \BibitemOpen
  \bibfield  {author} {\bibinfo {author} {\bibfnamefont {B.}~\bibnamefont
  {Dollet}}, \bibinfo {author} {\bibfnamefont {P.}~\bibnamefont {Marmottant}},
  \ and\ \bibinfo {author} {\bibfnamefont {V.}~\bibnamefont {Garbin}},\
  }\bibfield  {title} {\enquote {\bibinfo {title} {Bubble dynamics in soft and
  biological matter},}\ }\href@noop {} {\bibfield  {journal} {\bibinfo
  {journal} {Annual Review of Fluid Mechanics}\ }\textbf {\bibinfo {volume}
  {51}},\ \bibinfo {pages} {331--355} (\bibinfo {year} {2019})}\BibitemShut
  {NoStop}%
\bibitem [{\citenamefont {Jamburidze}\ \emph {et~al.}(2017)\citenamefont
  {Jamburidze}, \citenamefont {De~Corato}, \citenamefont {Huerre},
  \citenamefont {Pommella},\ and\ \citenamefont {Garbin}}]{jamburidze2017high}%
  \BibitemOpen
  \bibfield  {author} {\bibinfo {author} {\bibfnamefont {A.}~\bibnamefont
  {Jamburidze}}, \bibinfo {author} {\bibfnamefont {M.}~\bibnamefont
  {De~Corato}}, \bibinfo {author} {\bibfnamefont {A.}~\bibnamefont {Huerre}},
  \bibinfo {author} {\bibfnamefont {A.}~\bibnamefont {Pommella}}, \ and\
  \bibinfo {author} {\bibfnamefont {V.}~\bibnamefont {Garbin}},\ }\bibfield
  {title} {\enquote {\bibinfo {title} {High-frequency linear rheology of
  hydrogels probed by ultrasound-driven microbubble dynamics},}\ }\href@noop {}
  {\bibfield  {journal} {\bibinfo  {journal} {Soft Matter}\ }\textbf {\bibinfo
  {volume} {13}},\ \bibinfo {pages} {3946--3953} (\bibinfo {year}
  {2017})}\BibitemShut {NoStop}%
\bibitem [{\citenamefont {Novak}(2007)}]{novak2007molecular}%
  \BibitemOpen
  \bibfield  {author} {\bibinfo {author} {\bibfnamefont {B.~R.}\ \bibnamefont
  {Novak}},\ }\href@noop {} {\emph {\bibinfo {title} {Molecular simulation
  studies of heterogeneous bubble nucleation: effects of surface chemistry and
  topology}}}\ (\bibinfo  {publisher} {University of Notre Dame},\ \bibinfo
  {year} {2007})\BibitemShut {NoStop}%
\bibitem [{\citenamefont {Rayleigh}(1917)}]{rayleigh1917viii}%
  \BibitemOpen
  \bibfield  {author} {\bibinfo {author} {\bibfnamefont {L.}~\bibnamefont
  {Rayleigh}},\ }\bibfield  {title} {\enquote {\bibinfo {title} {Viii. on the
  pressure developed in a liquid during the collapse of a spherical cavity},}\
  }\href@noop {} {\bibfield  {journal} {\bibinfo  {journal} {The London,
  Edinburgh, and Dublin Philosophical Magazine and Journal of Science}\
  }\textbf {\bibinfo {volume} {34}},\ \bibinfo {pages} {94--98} (\bibinfo
  {year} {1917})}\BibitemShut {NoStop}%
\bibitem [{\citenamefont {Maruyama}, \citenamefont {Kimura},\ and\
  \citenamefont {Yamaguchi}(1997)}]{maruyama1997molecular}%
  \BibitemOpen
  \bibfield  {author} {\bibinfo {author} {\bibfnamefont {S.}~\bibnamefont
  {Maruyama}}, \bibinfo {author} {\bibfnamefont {T.}~\bibnamefont {Kimura}}, \
  and\ \bibinfo {author} {\bibfnamefont {Y.}~\bibnamefont {Yamaguchi}},\
  }\bibfield  {title} {\enquote {\bibinfo {title} {A molecular dynamics
  simulation of a bubble nucleation on solid surface},}\ }in\ \href@noop {}
  {\emph {\bibinfo {booktitle} {National Heat Transfer Symposium of Japan}}},\
  Vol.~\bibinfo {volume} {34}\ (\bibinfo {organization} {Citeseer},\ \bibinfo
  {year} {1997})\ pp.\ \bibinfo {pages} {675--676}\BibitemShut {NoStop}%
\bibitem [{\citenamefont {Li}, \citenamefont {Vlahovska},\ and\ \citenamefont
  {Karniadakis}(2013)}]{li2013continuum}%
  \BibitemOpen
  \bibfield  {author} {\bibinfo {author} {\bibfnamefont {X.}~\bibnamefont
  {Li}}, \bibinfo {author} {\bibfnamefont {P.~M.}\ \bibnamefont {Vlahovska}}, \
  and\ \bibinfo {author} {\bibfnamefont {G.~E.}\ \bibnamefont {Karniadakis}},\
  }\bibfield  {title} {\enquote {\bibinfo {title} {Continuum-and particle-based
  modeling of shapes and dynamics of red blood cells in health and disease},}\
  }\href@noop {} {\bibfield  {journal} {\bibinfo  {journal} {Soft Matter}\
  }\textbf {\bibinfo {volume} {9}},\ \bibinfo {pages} {28--37} (\bibinfo {year}
  {2013})}\BibitemShut {NoStop}%
\bibitem [{\citenamefont {Wang}\ \emph {et~al.}(2020)\citenamefont {Wang},
  \citenamefont {Li}, \citenamefont {Ouyang},\ and\ \citenamefont
  {Karniadakis}}]{wang2020controlled}%
  \BibitemOpen
  \bibfield  {author} {\bibinfo {author} {\bibfnamefont {Y.}~\bibnamefont
  {Wang}}, \bibinfo {author} {\bibfnamefont {Z.}~\bibnamefont {Li}}, \bibinfo
  {author} {\bibfnamefont {J.}~\bibnamefont {Ouyang}}, \ and\ \bibinfo {author}
  {\bibfnamefont {G.~E.}\ \bibnamefont {Karniadakis}},\ }\bibfield  {title}
  {\enquote {\bibinfo {title} {Controlled release of entrapped nanoparticles
  from thermoresponsive hydrogels with tunable network characteristics},}\
  }\href@noop {} {\bibfield  {journal} {\bibinfo  {journal} {Soft Matter}\
  }\textbf {\bibinfo {volume} {16}},\ \bibinfo {pages} {4756--4766} (\bibinfo
  {year} {2020})}\BibitemShut {NoStop}%
\bibitem [{\citenamefont {Li}\ \emph {et~al.}(2013)\citenamefont {Li},
  \citenamefont {Hu}, \citenamefont {Wang}, \citenamefont {Ma},\ and\
  \citenamefont {Zhou}}]{li2013three}%
  \BibitemOpen
  \bibfield  {author} {\bibinfo {author} {\bibfnamefont {Z.}~\bibnamefont
  {Li}}, \bibinfo {author} {\bibfnamefont {G.}~\bibnamefont {Hu}}, \bibinfo
  {author} {\bibfnamefont {Z.}~\bibnamefont {Wang}}, \bibinfo {author}
  {\bibfnamefont {Y.}~\bibnamefont {Ma}}, \ and\ \bibinfo {author}
  {\bibfnamefont {Z.}~\bibnamefont {Zhou}},\ }\bibfield  {title} {\enquote
  {\bibinfo {title} {Three dimensional flow structures in a moving droplet on
  substrate: A dissipative particle dynamics study},}\ }\href@noop {}
  {\bibfield  {journal} {\bibinfo  {journal} {Physics of Fluids}\ }\textbf
  {\bibinfo {volume} {25}},\ \bibinfo {pages} {072103} (\bibinfo {year}
  {2013})}\BibitemShut {NoStop}%
\bibitem [{\citenamefont {Zhang}\ \emph {et~al.}(2019)\citenamefont {Zhang},
  \citenamefont {Li}, \citenamefont {Maxey}, \citenamefont {Chen},\ and\
  \citenamefont {Karniadakis}}]{zhang2019self}%
  \BibitemOpen
  \bibfield  {author} {\bibinfo {author} {\bibfnamefont {K.}~\bibnamefont
  {Zhang}}, \bibinfo {author} {\bibfnamefont {Z.}~\bibnamefont {Li}}, \bibinfo
  {author} {\bibfnamefont {M.}~\bibnamefont {Maxey}}, \bibinfo {author}
  {\bibfnamefont {S.}~\bibnamefont {Chen}}, \ and\ \bibinfo {author}
  {\bibfnamefont {G.~E.}\ \bibnamefont {Karniadakis}},\ }\bibfield  {title}
  {\enquote {\bibinfo {title} {Self-cleaning of hydrophobic rough surfaces by
  coalescence-induced wetting transition},}\ }\href@noop {} {\bibfield
  {journal} {\bibinfo  {journal} {Langmuir}\ }\textbf {\bibinfo {volume}
  {35}},\ \bibinfo {pages} {2431--2442} (\bibinfo {year} {2019})}\BibitemShut
  {NoStop}%
\bibitem [{\citenamefont {Lin}\ \emph {et~al.}(2020)\citenamefont {Lin},
  \citenamefont {Yang}, \citenamefont {Chen}, \citenamefont {Chen},\ and\
  \citenamefont {Yin}}]{lin2020dissipative}%
  \BibitemOpen
  \bibfield  {author} {\bibinfo {author} {\bibfnamefont {C.}~\bibnamefont
  {Lin}}, \bibinfo {author} {\bibfnamefont {L.}~\bibnamefont {Yang}}, \bibinfo
  {author} {\bibfnamefont {F.}~\bibnamefont {Chen}}, \bibinfo {author}
  {\bibfnamefont {S.}~\bibnamefont {Chen}}, \ and\ \bibinfo {author}
  {\bibfnamefont {H.}~\bibnamefont {Yin}},\ }\bibfield  {title} {\enquote
  {\bibinfo {title} {A dissipative particle dynamics and discrete element
  method coupled model for particle interactions in sedimentation toward the
  fabrication of a functionally graded material},}\ }\href@noop {} {\bibfield
  {journal} {\bibinfo  {journal} {Colloids and Surfaces A: Physicochemical and
  Engineering Aspects}\ }\textbf {\bibinfo {volume} {604}},\ \bibinfo {pages}
  {125326} (\bibinfo {year} {2020})}\BibitemShut {NoStop}%
\bibitem [{\citenamefont {Wu}\ \emph {et~al.}(2017)\citenamefont {Wu},
  \citenamefont {Chu}, \citenamefont {Sheng},\ and\ \citenamefont
  {Tsao}}]{wu2017sliding}%
  \BibitemOpen
  \bibfield  {author} {\bibinfo {author} {\bibfnamefont {C.-J.}\ \bibnamefont
  {Wu}}, \bibinfo {author} {\bibfnamefont {K.-C.}\ \bibnamefont {Chu}},
  \bibinfo {author} {\bibfnamefont {Y.-J.}\ \bibnamefont {Sheng}}, \ and\
  \bibinfo {author} {\bibfnamefont {H.-K.}\ \bibnamefont {Tsao}},\ }\bibfield
  {title} {\enquote {\bibinfo {title} {Sliding dynamic behavior of a nanobubble
  on a surface},}\ }\href@noop {} {\bibfield  {journal} {\bibinfo  {journal}
  {The Journal of Physical Chemistry C}\ }\textbf {\bibinfo {volume} {121}},\
  \bibinfo {pages} {17932--17940} (\bibinfo {year} {2017})}\BibitemShut
  {NoStop}%
\bibitem [{\citenamefont {Tran-Duc}, \citenamefont {Phan-Thien},\ and\
  \citenamefont {Cheong~Khoo}(2013)}]{tran2013rheology}%
  \BibitemOpen
  \bibfield  {author} {\bibinfo {author} {\bibfnamefont {T.}~\bibnamefont
  {Tran-Duc}}, \bibinfo {author} {\bibfnamefont {N.}~\bibnamefont
  {Phan-Thien}}, \ and\ \bibinfo {author} {\bibfnamefont {B.}~\bibnamefont
  {Cheong~Khoo}},\ }\bibfield  {title} {\enquote {\bibinfo {title} {Rheology of
  bubble suspensions using dissipative particle dynamics. part i: A hard-core
  dpd particle model for gas bubbles},}\ }\href@noop {} {\bibfield  {journal}
  {\bibinfo  {journal} {Journal of Rheology}\ }\textbf {\bibinfo {volume}
  {57}},\ \bibinfo {pages} {1715--1737} (\bibinfo {year} {2013})}\BibitemShut
  {NoStop}%
\bibitem [{\citenamefont {Pan}\ \emph {et~al.}(2018)\citenamefont {Pan},
  \citenamefont {Zhao}, \citenamefont {Lin},\ and\ \citenamefont
  {Shao}}]{pan2018mesoscopic}%
  \BibitemOpen
  \bibfield  {author} {\bibinfo {author} {\bibfnamefont {D.}~\bibnamefont
  {Pan}}, \bibinfo {author} {\bibfnamefont {G.}~\bibnamefont {Zhao}}, \bibinfo
  {author} {\bibfnamefont {Y.}~\bibnamefont {Lin}}, \ and\ \bibinfo {author}
  {\bibfnamefont {X.}~\bibnamefont {Shao}},\ }\bibfield  {title} {\enquote
  {\bibinfo {title} {Mesoscopic modelling of microbubble in liquid with finite
  density ratio of gas to liquid},}\ }\href@noop {} {\bibfield  {journal}
  {\bibinfo  {journal} {EPL (Europhysics Letters)}\ }\textbf {\bibinfo {volume}
  {122}},\ \bibinfo {pages} {20003} (\bibinfo {year} {2018})}\BibitemShut
  {NoStop}%
\bibitem [{\citenamefont {Chen}\ and\ \citenamefont
  {Chen}(1995)}]{chen1995universal}%
  \BibitemOpen
  \bibfield  {author} {\bibinfo {author} {\bibfnamefont {T.}~\bibnamefont
  {Chen}}\ and\ \bibinfo {author} {\bibfnamefont {H.}~\bibnamefont {Chen}},\
  }\bibfield  {title} {\enquote {\bibinfo {title} {Universal approximation to
  nonlinear operators by neural networks with arbitrary activation functions
  and its application to dynamical systems},}\ }\href@noop {} {\bibfield
  {journal} {\bibinfo  {journal} {IEEE Transactions on Neural Networks}\
  }\textbf {\bibinfo {volume} {6}},\ \bibinfo {pages} {911--917} (\bibinfo
  {year} {1995})}\BibitemShut {NoStop}%
\bibitem [{\citenamefont {Lu}\ \emph {et~al.}(2019)\citenamefont {Lu},
  \citenamefont {Meng}, \citenamefont {Mao},\ and\ \citenamefont
  {Karniadakis}}]{lu2019deepxde}%
  \BibitemOpen
  \bibfield  {author} {\bibinfo {author} {\bibfnamefont {L.}~\bibnamefont
  {Lu}}, \bibinfo {author} {\bibfnamefont {X.}~\bibnamefont {Meng}}, \bibinfo
  {author} {\bibfnamefont {Z.}~\bibnamefont {Mao}}, \ and\ \bibinfo {author}
  {\bibfnamefont {G.~E.}\ \bibnamefont {Karniadakis}},\ }\bibfield  {title}
  {\enquote {\bibinfo {title} {{DeepXDE}: A deep learning library for solving
  differential equations},}\ }\href@noop {} {\bibfield  {journal} {\bibinfo
  {journal} {arXiv preprint arXiv:1907.04502}\ } (\bibinfo {year}
  {2019})}\BibitemShut {NoStop}%
\bibitem [{\citenamefont {Plesset}\ and\ \citenamefont
  {Prosperetti}(1977)}]{plesset1977bubble}%
  \BibitemOpen
  \bibfield  {author} {\bibinfo {author} {\bibfnamefont {M.~S.}\ \bibnamefont
  {Plesset}}\ and\ \bibinfo {author} {\bibfnamefont {A.}~\bibnamefont
  {Prosperetti}},\ }\bibfield  {title} {\enquote {\bibinfo {title} {Bubble
  dynamics and cavitation},}\ }\href@noop {} {\bibfield  {journal} {\bibinfo
  {journal} {Annual Review of Fluid Mechanics}\ }\textbf {\bibinfo {volume}
  {9}},\ \bibinfo {pages} {145--185} (\bibinfo {year} {1977})}\BibitemShut
  {NoStop}%
\bibitem [{\citenamefont {Brennen}(2013)}]{brennen2013cavitation}%
  \BibitemOpen
  \bibfield  {author} {\bibinfo {author} {\bibfnamefont {C.~E.}\ \bibnamefont
  {Brennen}},\ }\href@noop {} {\emph {\bibinfo {title} {Cavitation and bubble
  dynamics}}}\ (\bibinfo  {publisher} {Cambridge University Press},\ \bibinfo
  {year} {2013})\BibitemShut {NoStop}%
\bibitem [{\citenamefont {Prosperetti}(2017)}]{prosperetti2017vapor}%
  \BibitemOpen
  \bibfield  {author} {\bibinfo {author} {\bibfnamefont {A.}~\bibnamefont
  {Prosperetti}},\ }\bibfield  {title} {\enquote {\bibinfo {title} {Vapor
  bubbles},}\ }\href@noop {} {\bibfield  {journal} {\bibinfo  {journal} {Annual
  Review of Fluid Mechanics}\ }\textbf {\bibinfo {volume} {49}} (\bibinfo
  {year} {2017})}\BibitemShut {NoStop}%
\bibitem [{\citenamefont {Hochreiter}\ and\ \citenamefont
  {Schmidhuber}(1997)}]{hochreiter1997long}%
  \BibitemOpen
  \bibfield  {author} {\bibinfo {author} {\bibfnamefont {S.}~\bibnamefont
  {Hochreiter}}\ and\ \bibinfo {author} {\bibfnamefont {J.}~\bibnamefont
  {Schmidhuber}},\ }\bibfield  {title} {\enquote {\bibinfo {title} {Long
  short-term memory},}\ }\href@noop {} {\bibfield  {journal} {\bibinfo
  {journal} {Neural Computation}\ }\textbf {\bibinfo {volume} {9}},\ \bibinfo
  {pages} {1735--1780} (\bibinfo {year} {1997})}\BibitemShut {NoStop}%
\bibitem [{\citenamefont {Fu}, \citenamefont {Zhang},\ and\ \citenamefont
  {Li}(2016)}]{fu2016using}%
  \BibitemOpen
  \bibfield  {author} {\bibinfo {author} {\bibfnamefont {R.}~\bibnamefont
  {Fu}}, \bibinfo {author} {\bibfnamefont {Z.}~\bibnamefont {Zhang}}, \ and\
  \bibinfo {author} {\bibfnamefont {L.}~\bibnamefont {Li}},\ }\bibfield
  {title} {\enquote {\bibinfo {title} {Using {LSTM} and {GRU} neural network
  methods for traffic flow prediction},}\ }in\ \href@noop {} {\emph {\bibinfo
  {booktitle} {2016 31st Youth Academic Annual Conference of Chinese
  Association of Automation (YAC)}}}\ (\bibinfo {organization} {IEEE},\
  \bibinfo {year} {2016})\ pp.\ \bibinfo {pages} {324--328}\BibitemShut
  {NoStop}%
\bibitem [{\citenamefont {del {\'A}guila~Ferrandis}\ \emph
  {et~al.}(2019)\citenamefont {del {\'A}guila~Ferrandis}, \citenamefont
  {Triantafyllou}, \citenamefont {Chryssostomidis},\ and\ \citenamefont
  {Karniadakis}}]{del2019learning}%
  \BibitemOpen
  \bibfield  {author} {\bibinfo {author} {\bibfnamefont {J.}~\bibnamefont {del
  {\'A}guila~Ferrandis}}, \bibinfo {author} {\bibfnamefont {M.}~\bibnamefont
  {Triantafyllou}}, \bibinfo {author} {\bibfnamefont {C.}~\bibnamefont
  {Chryssostomidis}}, \ and\ \bibinfo {author} {\bibfnamefont {G.~E.}\
  \bibnamefont {Karniadakis}},\ }\bibfield  {title} {\enquote {\bibinfo {title}
  {Learning functionals via {LSTM} neural networks for predicting vessel
  dynamics in extreme sea states.}}\ }\href@noop {} {\bibfield  {journal}
  {\bibinfo  {journal} {arXiv preprint arXiv:1912.13382}\ } (\bibinfo {year}
  {2019})}\BibitemShut {NoStop}%
\bibitem [{\citenamefont {Alahi}\ \emph {et~al.}(2016)\citenamefont {Alahi},
  \citenamefont {Goel}, \citenamefont {Ramanathan}, \citenamefont {Robicquet},
  \citenamefont {Feifei},\ and\ \citenamefont {Savarese}}]{alahi2016social}%
  \BibitemOpen
  \bibfield  {author} {\bibinfo {author} {\bibfnamefont {A.}~\bibnamefont
  {Alahi}}, \bibinfo {author} {\bibfnamefont {K.}~\bibnamefont {Goel}},
  \bibinfo {author} {\bibfnamefont {V.}~\bibnamefont {Ramanathan}}, \bibinfo
  {author} {\bibfnamefont {A.}~\bibnamefont {Robicquet}}, \bibinfo {author}
  {\bibfnamefont {L.}~\bibnamefont {Feifei}}, \ and\ \bibinfo {author}
  {\bibfnamefont {S.}~\bibnamefont {Savarese}},\ }\bibfield  {title} {\enquote
  {\bibinfo {title} {Social {LSTM}: Human trajectory prediction in crowded
  spaces},}\ }in\ \href@noop {} {\emph {\bibinfo {booktitle} {Proceedings of
  the IEEE conference on computer vision and pattern recognition}}}\ (\bibinfo
  {year} {2016})\ pp.\ \bibinfo {pages} {961--971}\BibitemShut {NoStop}%
\bibitem [{\citenamefont {Nelson}, \citenamefont {Pereira},\ and\ \citenamefont
  {de~Oliveira}(2017)}]{nelson2017stock}%
  \BibitemOpen
  \bibfield  {author} {\bibinfo {author} {\bibfnamefont {D.~M.}\ \bibnamefont
  {Nelson}}, \bibinfo {author} {\bibfnamefont {A.~C.}\ \bibnamefont {Pereira}},
  \ and\ \bibinfo {author} {\bibfnamefont {R.~A.}\ \bibnamefont
  {de~Oliveira}},\ }\bibfield  {title} {\enquote {\bibinfo {title} {Stock
  market's price movement prediction with {LSTM} neural networks},}\ }in\
  \href@noop {} {\emph {\bibinfo {booktitle} {2017 International joint
  conference on neural networks (IJCNN)}}}\ (\bibinfo {organization} {IEEE},\
  \bibinfo {year} {2017})\ pp.\ \bibinfo {pages} {1419--1426}\BibitemShut
  {NoStop}%
\bibitem [{\citenamefont {Groot}\ and\ \citenamefont
  {Warren}(1997)}]{groot1997dissipative}%
  \BibitemOpen
  \bibfield  {author} {\bibinfo {author} {\bibfnamefont {R.~D.}\ \bibnamefont
  {Groot}}\ and\ \bibinfo {author} {\bibfnamefont {P.~B.}\ \bibnamefont
  {Warren}},\ }\bibfield  {title} {\enquote {\bibinfo {title} {Dissipative
  particle dynamics: Bridging the gap between atomistic and mesoscopic
  simulation},}\ }\href@noop {} {\bibfield  {journal} {\bibinfo  {journal} {The
  Journal of Chemical Physics}\ }\textbf {\bibinfo {volume} {107}},\ \bibinfo
  {pages} {4423--4435} (\bibinfo {year} {1997})}\BibitemShut {NoStop}%
\bibitem [{\citenamefont {Espa{\~n}ol}\ and\ \citenamefont
  {Warren}(1995)}]{espanol1995statistical}%
  \BibitemOpen
  \bibfield  {author} {\bibinfo {author} {\bibfnamefont {P.}~\bibnamefont
  {Espa{\~n}ol}}\ and\ \bibinfo {author} {\bibfnamefont {P.}~\bibnamefont
  {Warren}},\ }\bibfield  {title} {\enquote {\bibinfo {title} {Statistical
  mechanics of dissipative particle dynamics},}\ }\href@noop {} {\bibfield
  {journal} {\bibinfo  {journal} {EPL (Europhysics Letters)}\ }\textbf
  {\bibinfo {volume} {30}},\ \bibinfo {pages} {191} (\bibinfo {year}
  {1995})}\BibitemShut {NoStop}%
\bibitem [{\citenamefont {Warren}(2003)}]{warren2003vapor}%
  \BibitemOpen
  \bibfield  {author} {\bibinfo {author} {\bibfnamefont {P.}~\bibnamefont
  {Warren}},\ }\bibfield  {title} {\enquote {\bibinfo {title} {Vapor-liquid
  coexistence in many-body dissipative particle dynamics},}\ }\href@noop {}
  {\bibfield  {journal} {\bibinfo  {journal} {Physical Review E}\ }\textbf
  {\bibinfo {volume} {68}},\ \bibinfo {pages} {066702} (\bibinfo {year}
  {2003})}\BibitemShut {NoStop}%
\bibitem [{\citenamefont {Arienti}\ \emph {et~al.}(2011)\citenamefont
  {Arienti}, \citenamefont {Pan}, \citenamefont {Li},\ and\ \citenamefont
  {Karniadakis}}]{arienti2011many}%
  \BibitemOpen
  \bibfield  {author} {\bibinfo {author} {\bibfnamefont {M.}~\bibnamefont
  {Arienti}}, \bibinfo {author} {\bibfnamefont {W.}~\bibnamefont {Pan}},
  \bibinfo {author} {\bibfnamefont {X.}~\bibnamefont {Li}}, \ and\ \bibinfo
  {author} {\bibfnamefont {G.~E.}\ \bibnamefont {Karniadakis}},\ }\bibfield
  {title} {\enquote {\bibinfo {title} {Many-body dissipative particle dynamics
  simulation of liquid/vapor and liquid/solid interactions},}\ }\href@noop {}
  {\bibfield  {journal} {\bibinfo  {journal} {The Journal of Chemical Physics}\
  }\textbf {\bibinfo {volume} {134}},\ \bibinfo {pages} {204114} (\bibinfo
  {year} {2011})}\BibitemShut {NoStop}%
\bibitem [{\citenamefont {Pagonabarraga}\ and\ \citenamefont
  {Frenkel}(2001)}]{pagonabarraga2001dissipative}%
  \BibitemOpen
  \bibfield  {author} {\bibinfo {author} {\bibfnamefont {I.}~\bibnamefont
  {Pagonabarraga}}\ and\ \bibinfo {author} {\bibfnamefont {D.}~\bibnamefont
  {Frenkel}},\ }\bibfield  {title} {\enquote {\bibinfo {title} {Dissipative
  particle dynamics for interacting systems},}\ }\href@noop {} {\bibfield
  {journal} {\bibinfo  {journal} {The Journal of Chemical Physics}\ }\textbf
  {\bibinfo {volume} {115}},\ \bibinfo {pages} {5015--5026} (\bibinfo {year}
  {2001})}\BibitemShut {NoStop}%
\bibitem [{\citenamefont {Li}\ \emph {et~al.}(2018)\citenamefont {Li},
  \citenamefont {Bian}, \citenamefont {Tang},\ and\ \citenamefont
  {Karniadakis}}]{li2018dissipative}%
  \BibitemOpen
  \bibfield  {author} {\bibinfo {author} {\bibfnamefont {Z.}~\bibnamefont
  {Li}}, \bibinfo {author} {\bibfnamefont {X.}~\bibnamefont {Bian}}, \bibinfo
  {author} {\bibfnamefont {Y.}~\bibnamefont {Tang}}, \ and\ \bibinfo {author}
  {\bibfnamefont {G.~E.}\ \bibnamefont {Karniadakis}},\ }\bibfield  {title}
  {\enquote {\bibinfo {title} {A dissipative particle dynamics method for
  arbitrarily complex geometries},}\ }\href@noop {} {\bibfield  {journal}
  {\bibinfo  {journal} {Journal of Computational Physics}\ }\textbf {\bibinfo
  {volume} {355}},\ \bibinfo {pages} {534--547} (\bibinfo {year}
  {2018})}\BibitemShut {NoStop}%
\bibitem [{\citenamefont {Rycroft}(2009)}]{rycroft2009voro}%
  \BibitemOpen
  \bibfield  {author} {\bibinfo {author} {\bibfnamefont {C.~H.}\ \bibnamefont
  {Rycroft}},\ }\bibfield  {title} {\enquote {\bibinfo {title} {{VORO++}: a
  three-dimensional voronoi cell library in {C++}.}}\ }\href@noop {} {\bibfield
   {journal} {\bibinfo  {journal} {Chaos}\ }\textbf {\bibinfo {volume} {19}},\
  \bibinfo {pages} {041111} (\bibinfo {year} {2009})}\BibitemShut {NoStop}%
\bibitem [{\citenamefont {Tsai}(1979)}]{tsai1979virial}%
  \BibitemOpen
  \bibfield  {author} {\bibinfo {author} {\bibfnamefont {D.}~\bibnamefont
  {Tsai}},\ }\bibfield  {title} {\enquote {\bibinfo {title} {The virial theorem
  and stress calculation in molecular dynamics},}\ }\href@noop {} {\bibfield
  {journal} {\bibinfo  {journal} {The Journal of Chemical Physics}\ }\textbf
  {\bibinfo {volume} {70}},\ \bibinfo {pages} {1375--1382} (\bibinfo {year}
  {1979})}\BibitemShut {NoStop}%
\bibitem [{\citenamefont {Lin}\ \emph {et~al.}(2018)\citenamefont {Lin},
  \citenamefont {Chen}, \citenamefont {Xiao},\ and\ \citenamefont
  {Liu}}]{lin2018tuning}%
  \BibitemOpen
  \bibfield  {author} {\bibinfo {author} {\bibfnamefont {C.}~\bibnamefont
  {Lin}}, \bibinfo {author} {\bibfnamefont {S.}~\bibnamefont {Chen}}, \bibinfo
  {author} {\bibfnamefont {L.}~\bibnamefont {Xiao}}, \ and\ \bibinfo {author}
  {\bibfnamefont {Y.}~\bibnamefont {Liu}},\ }\bibfield  {title} {\enquote
  {\bibinfo {title} {Tuning drop motion by chemical chessboard-patterned
  surfaces: a many-body dissipative particle dynamics study},}\ }\href@noop {}
  {\bibfield  {journal} {\bibinfo  {journal} {Langmuir}\ }\textbf {\bibinfo
  {volume} {34}},\ \bibinfo {pages} {2708--2715} (\bibinfo {year}
  {2018})}\BibitemShut {NoStop}%
\end{thebibliography}%

\end{document}